\documentclass[secnumeq]{FBSart2}
\usepackage{amsfonts}
\usepackage{amssymb}
\usepackage{amsmath}
\usepackage{epsfig,bm}
\usepackage{slashbox}
\usepackage{color}

\newcommand{\vi}[1]{\mbox{\boldmath $#1$}}

\title{
Four-nucleon scattering with a correlated Gaussian basis method}
\author{S. Aoyama\instnr{1}, K. Arai\instnr{2},
Y. Suzuki\instnr{3,4},
P. Descouvemont\instnr{5},
D. Baye\instnr{6,5}}

\instlist{Center for Academic Information Service,
Niigata University, Niigata 950-2181, Japan
\and 
Division of General Education, Nagaoka National College of
Technology,
888 Nishikatakai, Nagaoka,Niigata, 940-8532, Japan
\and
Department of Physics, Niigata University, Niigata 950-2181, Japan
\and
RIKEN Nishina Center, RIKEN, Wako 351-0198, Japan
\and Physique Nucl\'eaire Th\'eorique et Physique Math\'ematique, C.P. 229,
Universit\'e Libre de Bruxelles (ULB), B 1050 Brussels, Belgium
\and Physique Quantique, C.P. 165/82,  
Universit\'e Libre de Bruxelles (ULB), B 1050 Brussels, Belgium
}

\runningauthor{S. Aoyama, K. Arai, Y. Suzuki, P. Descouvemont, D. Baye}
\runningtitle{
Four nucleon scattering with correlated Gaussian basis method}
\begin{document}
\maketitle

\begin{abstract}
Elastic-scattering phase shifts for four-nucleon systems
are studied in an $ab$-$initio$ type cluster model
in order to clarify the role of the tensor force and to 
investigate cluster distortions
in low energy $d$+$d$ and $t$+$p$ scattering. 
In the present method, the description of the cluster wave function
is extended from a simple (0$s$) harmonic-oscillator shell model
to a few-body model with a realistic interaction, in which
the wave function of the subsystems are determined with
the Stochastic Variational Method.
In order to calculate the matrix elements
of the four-body system,
we have developed a Triple Global Vector Representation method
for the correlated Gaussian basis functions.
To compare effects of the cluster distortion with realistic and effective interactions,
we employ the AV8$^{\prime}$ potential as a realistic interaction and
the Minnesota potential as an effective interaction.
Especially for $^1S_0$, the calculated phase shifts
show that the 
$t$+$p$ and $h$+$n$ channels are strongly coupled to the $d$+$d$ channel
for the case of the realistic interaction.
On the contrary, the coupling of these channels plays a relatively
minor role for the case of the effective interaction.
This difference between both potentials originates from the tensor term in the realistic interaction.
Furthermore, the tensor interaction makes the energy splitting
of the negative parity states of $^4$He consistent with experiments.
No such splitting is however reproduced with the effective interaction. 
\end{abstract}

\section{Introduction}	
\label{sect.1}

The microscopic cluster model is one of the successful models
to study the structure and reactions of light nuclei \cite{tang77}.
In the conventional cluster model, one assumes that the nucleus is composed of
several simple clusters with $A \le 4$ which are described by (0$s$) harmonic-oscillator shell model
functions,
and use an effective $N$-$N$ interaction which is appropriate for
such a model space.
However, it is well known that the ground states of the typical clusters 
$d$, $t$, $h$ and $^4$He have non-negligible admixtures
of $D$-wave component due to the tensor interaction.
Since the conventional cluster model does not directly treat the $D$-wave
component, the strong attraction of the nucleon-nucleon interaction
due to the tensor term is assumed to be 
renormalized into the central term
of the effective interaction.

Recently, $ab$-$initio$ structure calculations \cite{benchmark}
have been successfully developed:  
Stochastic Variational Method (SVM) \cite{varga94,varga97,book,vs95},
Global Vector Representation method (GVR) \cite{DGVR,GVR},
Green's function Monte Carlo method \cite{carlson98},
no core shell model \cite{navratil00}, correlated hyperspherical harmonics
method \cite{viviani98r},
unitary correlation operator method \cite{ucom}, and so on.
Although the application of $ab$-$initio$ reaction calculations 
with a realistic interaction
are restricted so much in comparison with structure calculations,
it has been intensively applied to the four-nucleon systems
$t$+$n$ and $h$+$p$ \cite{arai10,hofmann01,deltuva07,navratil09,viviani98,lazauskas05,fisher06}.
Especially $d$+$d$ scattering states, which couple
to  $t$+$p$ and $h$+$n$ channels, have
attracted much attention, because the $d$+$d$ radiative capture
is one of the mechanisms
making $^4$He through electro-magnetic transitions \cite{arriaga91,sabourov04} 
and also have posed intriguing puzzles for analyzing powers 
\cite{hofmann08,hofmann97,deltuva08,lazauskas04,ciesielski99}, which are motivated by
the famous $A_y$ problem in the three-nucleon system.

Furthermore, the $d$+$d$ elastic-scattering phase shifts are interesting
because the astrophysical S-factor of the $d$($d$,$\gamma$)$^4$He reaction is not
explained by any calculation using an effective interaction
that contains no tensor term,
and is expected to be contributed by the $D$-wave components of 
the clusters
through $E2$ transitions \cite{langanke87,fowler67}.

Also, thanks to recent developments of  the microscopic cluster model,
the simple model using the (0$s$)
harmonic-oscillator wave function 
with  an effective interaction is not mandatory any more, at least,
in light nuclei.
We can use a kind of $ab$-$initio$ cluster model
which employs more realistic cluster wave functions with realistic interactions.
Therefore, it is interesting to see the difference
between the $ab$-$initio$ reaction calculations with a tensor term
and the conventional
cluster model calculations without a tensor term in few-body systems.

The microscopic $R$-matrix method (MRM) with a cluster model (GCM or RGM)
has been applied to studies of many nuclei \cite{baye77,kanada85,arai01,desc10}.
It is now used in $ab$ $initio$ descriptions of collisions \cite{navratil09}. 
We have also applied the MRM to the $h$+$p$ scattering problem with
more realistic cluster wave functions by using a realistic interaction \cite{arai10}.
The Gaussian basis functions for the expansion of the cluster wave functions 
are chosen by a technique of the SVM \cite{book}.
In the MRM, as will be shown later, the relative wave function
between clusters ($a$ and $b$) is connected to the boundary condition
at a channel radius.
The problem is how to calculate the matrix elements.
In this paper, we develop a method called
the Triple Global Vector Representation method (TGVR),
by which we calculate the matrix elements in a unified way.
Although we restricted ourselves to four
nucleon systems in the present paper, the formulation of the TGVR itself
can be applied to more than four-body systems as in the previous studies
of the Global Vector Representation methods (GVR) \cite{GVR}.
Furthermore, for scattering problems, the TGVR 
can deal with more complicated systems than 
the double (or single) global vector
which was given in the previous papers \cite{DGVR,GVR}, because we need
three  representative orbital angular momenta, 
the total internal orbital momenta of both clusters and 
the orbital momentum of their relative motion, 
in order to reasonably describe
the scattering states.
In other words,
the first global vector represents the angular momentum
of cluster $a$, the second  global vector represents the angular momentum
of cluster $b$, and the third global vector represents the relative
angular momentum between the clusters.

In this paper, we will investigate the effect of the distortion of clusters
on the $d$+$d$ elastic-scattering by comparing 
the phase shifts calculated with 
a realistic and an effective interaction.
In section 2, we explain the MRM in brief.
In section 3, 
the correlated Gaussian (CG) method with the TGVR, which has newly been
developed for the present analysis, will be presented.
In section 4, we will explain how to calculate the matrix elements
with TGVR basis functions. The typical matrix elements are also given in the
appendix.
In section 5, we will present and discuss the calculated 
scattering phase shifts in detail.
Finally, summary and conclusions are given in section 6.

\section{Microscopic $R$-matrix method}
\label{sect.2}
In the present study we calculate $d$+$d$ and $t$+$p$ (and $h$+$n$) elastic
scattering phase shifts with 
the microscopic $R$-matrix method. Though the method is well documented in
e.g. Refs.~\cite{baye77,kanada85,desc10}, we briefly explain it below in order to present 
definitions and equations needed in the subsequent sections.
Since our interest is on low-energy scattering, we consider only
two-body channels. A channel $\alpha$ is specified by the two nuclei 
(clusters) $a,
b$, their angular momenta, $I_a, I_b$, the channel spin $I$ 
that is a resultant of the 
coupling of $I_a$ and $I_b$, and the orbital angular momentum $\ell$
for the relative motion of $a$ and $b$.  
The wave function of channel $\alpha$ with the total angular momentum
$J$, its projection $M$, and the parity $\pi$ takes the form 
\begin{eqnarray}
\Psi^{JM\pi}_{\alpha}=
{\cal A}
\left[\left[\Phi^{a}_{I_a}\Phi^{b}_{I_b}\right]_I  
\chi_{\alpha}
(\mbox{\boldmath$\rho$}_{\alpha})\right]_{JM},
\label{wf-ch1}
\end{eqnarray}
where $\Phi^{a}_{I_a}$ and $\Phi^{b}_{I_b}$ are  respectively antisymmetrized 
intrinsic wave functions of $a$ and $b$, and 
${\cal A}$ is an operator that antisymmetrizes between the clusters.  
The square bracket $[{I_a} \ {I_b}]_I$ denotes the angular 
momentum coupling.
The coordinate ${\vi \rho}_\alpha$ in the relative motion function 
$\chi^J_{\alpha}$ is the relative distance vector of the clusters. 
The channel spin $I$ and the relative angular momentum $\ell$ in $\alpha$
are coupled to give the total angular momentum $J$.
The relative-motion functions $\chi_{\alpha}$ also depend on $J$ and $\pi$. 
For simplicity, this dependence is not displayed explicitly in the notation
for $\chi_{\alpha}$ as well as for some other quantities below. 

The configuration space is divided into two regions, 
internal and external, by the channel radius $a$. 
In the internal region ($\rho_\alpha\le a$), the total wave function 
may be expressed in terms of a combination of various
$\Psi_{\alpha}^{JM\pi}$s 
\begin{eqnarray}
\Psi^{JM\pi}_{\rm int}&=&\sum_{\alpha} \Psi^{JM\pi}_{\alpha}\nonumber \\
&=&\sum_{\alpha}\sum_{n}f_{\alpha n} {\cal A} 
u_{\alpha n}(\rho_\alpha) \phi^{JM\pi}_{\alpha},
\label{wf.int}
\end{eqnarray}
with
\begin{eqnarray}
\phi^{JM\pi}_{\alpha} = \frac{1}{\sqrt{(1+\delta_{I_aI_b}\delta_{ab})(1+\delta_{ab})}} \left\{
\left[\left[\Phi^{a}_{I_a}\Phi^{b}_{I_b}\right]_I Y_{\ell}(\widehat{\vi \rho}_\alpha)\right]_{JM} 
\right. \nonumber \\ \left.
+(-1)^{A_a+I_a+I_b-I+\ell}
\left[\left[\Phi^{b}_{I_b}\Phi^{a}_{I_a}\right]_I Y_{\ell}(\widehat{\vi \rho}_\alpha)\right]_{JM} \delta_{ab}
\right\},
\label{wf.channel}
\end{eqnarray}
where $A_a$ is the number of nucleons in cluster $a$,  
$\delta_{ab}$ is unity if $a$ and $b$ are identical clusters and zero otherwise, 
and $\delta_{I_aI_b}$ is unity if the clusters are in identical states and zero otherwise. 
In the second line of Eq.~(\ref{wf.int}),  
the relative motion functions of Eq.~(\ref{wf-ch1}) are 
expanded in terms of some basis functions as  
\begin{eqnarray}
\chi_{\alpha m}({\vi \rho}_\alpha) = \sum_n  f_{\alpha n} 
u_{\alpha n}(\rho_\alpha)Y_{\ell m}(\widehat{\vi \rho}_\alpha).
\label{chi.expansion}
\end{eqnarray}
In what follows we take 
\begin{eqnarray}
u_{\alpha n}(\rho_\alpha)=\rho_\alpha^{\ell}\exp(-\frac{1}{2}\lambda_n \rho_\alpha^2)
\label{u.functions}
\end{eqnarray}
with a suitable set of $\lambda_n$s. 

In the external region ($\rho_\alpha\ge a$), the total wave function takes the
form 
\begin{eqnarray}
\Psi^{JM\pi}_{\rm ext}=\sum_{\alpha} g_{\alpha}(\rho_\alpha)
 \phi^{JM\pi}_{\alpha}
\label{wf.ext}.
\end{eqnarray}
Note that the antisymmetrization between the clusters 
is dropped in the external region under the condition that the channel radius
$a$ is large enough. 
The function $g_{\alpha}(\rho_\alpha)$ of Eq.~(\ref{wf.ext}) is a solution of 
the equation 
\begin{eqnarray}
\left[-\frac{\hbar^2}{2\mu_{\alpha}}\left( \frac{d^2}{d\rho_\alpha^2}
+\frac{2}{\rho_\alpha}\,\frac{d}{d\rho_\alpha}-\frac{\ell(\ell+1)}{\rho_\alpha^2} \right)+\frac{Z_aZ_be^2}{\rho_\alpha}\right] g_{\alpha}(\rho_\alpha) =
E_{\alpha} g_{\alpha}(\rho_\alpha),
\end{eqnarray}
where $\mu_{\alpha}$ is the reduced mass for the relative motion in
channel $\alpha$, $Z_ae$ and $Z_be$ are the charges of $a$ and $b$, and 
$E_{\alpha}=E-E_a-E_b$ is the energy for the relative motion, 
where $E$ is the total energy, and $E_a$ and $E_b$ are 
the internal energies for the clusters $a$ and $b$, respectively. 
For the scattering initiated through the channel $\alpha_0$, 
the asymptotic form of $g_{\alpha}$ for the open channel $\alpha$
$(E_{\alpha} \geq 0)$ is 
\begin{eqnarray}
g_{\alpha}(\rho_\alpha) = v_\alpha^{-1/2} \rho_\alpha^{-1} 
[I_{\alpha}(k_\alpha \rho_\alpha)\delta_{\alpha\,
 \alpha_0}-S_{\alpha \, \alpha_0}^{J\pi} O_{\alpha}(k_\alpha \rho_\alpha)],
\end{eqnarray}
where $k_\alpha=\sqrt{2\mu_\alpha |E_\alpha |}/\hbar$, $v_\alpha=\hbar k_\alpha/\mu_\alpha$ and 
$S_{\alpha \, \alpha_0}^{J\pi}$ is an element of the 
$S$-matrix (or collision matrix) to be determined. 
Here $I_{\alpha}(k_\alpha \rho_\alpha)$ and $O_{\alpha}(k_\alpha \rho_\alpha)$ are the 
incoming and outgoing waves defined by 
\begin{eqnarray}
I_{\alpha}(k_\alpha \rho_\alpha)=O_{\alpha}(k_\alpha \rho_\alpha)^*=
G_\ell(\eta_{\alpha}, k_\alpha \rho_\alpha)-iF_\ell(\eta_{\alpha}, k_\alpha \rho_\alpha),
\end{eqnarray}
with the regular and irregular Coulomb functions $F_\ell$ and $G_\ell$. 
The Sommerfeld parameter $\eta_{\alpha}$ is $\mu_\alpha Z_a Z_b e^2/\hbar^2
k_\alpha$. 
For a closed channel $\alpha$ $(E_\alpha < 0)$, 
the asymptotic form of $g_\alpha$ is given by the Whittaker function 
\begin{eqnarray}
g_{\alpha}(\rho_\alpha) \propto \rho_\alpha^{-1} 
W_{-\eta_{\alpha},\ell+1/2}(2k_\alpha\rho_\alpha).
\end{eqnarray}

The matrix elements $S_{\alpha \alpha_0}^{J\pi}$ are determined by solving 
a Schr\"odinger equation with a microscopic Hamiltonian $H$ involving the $A_a + A_b$ nucleons,
\begin{eqnarray}
(H+{\cal L} -E)\Psi_{\rm int}^{JM\pi}={\cal L}\Psi_{\rm ext}^{JM\pi},
\end{eqnarray}
with the Bloch operator ${\cal L}$ 
\begin{eqnarray}
{\cal L}=\sum_{\alpha}\frac{\hbar^2}{2\mu_\alpha a}
\vert \phi_{\alpha}^{JM \pi} \rangle \delta(\rho_\alpha-a)
\left(\frac{\partial}{\partial \rho_\alpha}-\frac{b_\alpha}{\rho_\alpha} \right)\rho_\alpha
\langle \phi_{\alpha}^{JM\pi}\vert,
\label{Bloch}
\end{eqnarray}
where the channel radius $a$ is chosen to be the same for all channels, 
and the $b_\alpha$ are arbitrary constants. Here, we choose $b_\alpha=0$
for the open channels and 
$b_\alpha=2k_\alpha a W'_{-\eta_\alpha,\ell+1/2}(2k_\alpha a)/W_{-\eta_\alpha,\ell+1/2}(2k_\alpha a)$
for the closed channels. 
The results do not depend on the choices for $b_\alpha$ but these values simplify the calculations. 
Notice that the projector on $|\phi_{\alpha}^{JM \pi}\rangle$ in Eq.~(\ref{Bloch}) is not essential 
in a microscopic calculation and can be dropped since the various channels are orthogonal 
at the channel radius.

The Bloch operator ensures that the logarithmic derivative of the wave function 
is continuous at the channel radius. 
In addition, 
$\Psi_{\rm int}^{JM\pi}$ must be equal to 
$\Psi_{\rm ext}^{JM\pi}$ at $\rho_\alpha=a$. 
Projecting the Schr\"odinger equation on a basis state, one obtains 
\begin{eqnarray}
\sum_{\alpha n} C_{\alpha' n', \alpha n} \, f_{\alpha n} 
= \langle \Phi_{\alpha' n'}^{JM\pi} \vert {\cal L} \vert \Psi_{\rm ext}^{JM\pi} \rangle
\label{SEq}
\end{eqnarray}
with 
\begin{eqnarray}
C_{\alpha' n', \alpha n} = 
\langle \Phi_{\alpha' n'}^{JM\pi} \vert H+{\cal L} -E 
\vert {\cal A}\Phi_{\alpha n}^{JM\pi} \rangle_{\rm int},
\end{eqnarray}
and 
\begin{equation}
\Phi_{\alpha n}^{JM\pi} = u_{\alpha n}(\rho_\alpha)\phi_{\alpha}^{JM\pi}.
\end{equation}
Here $\langle\vert {\cal O} \vert\rangle_{\rm int}$ indicates that the integration 
with respect to $\rho_\alpha$ is to be carried out in the internal region. Actually 
$\langle\vert {\cal O} \vert\rangle_{\rm int}$ is obtained by calculating the matrix element 
$\langle\vert {\cal O} \vert\rangle$ in the entire space and subtracting the 
corresponding external matrix element $\langle\vert {\cal O} \vert\rangle_{\rm ext}$
that is easily obtained because no intercluster 
antisymmetrization is needed. 
The $R$-matrix and $Z$-matrix are defined by 
\begin{eqnarray}
&&{\cal R}_{\alpha' \alpha} \equiv \frac{\hbar^2 a}{2}
\left(\frac{k_{\alpha'}}{\mu_{\alpha'}\mu_{\alpha}k_\alpha}\right)^\frac{1}{2}
\sum_{n' n}u_{\alpha' n'}(a)(C^{-1})_{\alpha' n', \alpha n}u_{\alpha n}(a),
\\
&&{\cal Z}_{\alpha' \alpha} \equiv I_{\alpha} (k_\alpha a) \delta_{\alpha' \alpha}-
{\cal R}_{\alpha' \alpha}k_\alpha a I'_{\alpha}(k_\alpha a).
\end{eqnarray}
The $S$-matrix is finally obtained as 
\begin{eqnarray}
S^{J\pi}=({\cal Z}^*)^{-1}{\cal Z}.
\end{eqnarray}
In this paper we focus on the elastic phase shifts  $\delta_{\alpha}^{J\pi}$
that are defined by the diagonal elements of the $S$-matrix,  
\begin{eqnarray}
S_{\alpha \alpha}^{J\pi}=\eta_\alpha^{J\pi} e^{2i\delta_\alpha^{J\pi}}.
\end{eqnarray}

We study four-nucleon scattering  involving the 
$d$+$d$, $t$+$p$ and $h$+$n$ channels in the energy region 
around and below the $d$+$d$ threshold. In Table~\ref{chan0} we list all 
possible  labels $^{2I+1}\ell_J$ of physical channels for $J^{\pi}=0^{\pm}$,
$1^{\pm}$, and $2^{\pm}$, assuming $\ell \leq 2$. 
Here ``physical'' means that the channels involve the cluster bound states 
that appear in the external region as well. 
Non-physical channels involving excited pseudo states 
will also be included in most calculations. 
Note that the $d$+$d$ channel must satisfy the condition of 
$I+\ell$ even (see Eq.~(\ref{wf.channel})).  The 
channel spin $I=0$ or 2 
can couple with only even $\ell$, but $I=1$ with only odd
$\ell$. It is noted that the relative motion for the $d$+$d$ 
scattering can have $\ell=0$ only when $J^{\pi}$ is equal to 
$0^+$ and $2^+$.

\begin{table}
\caption{Channel spins ($^{2I+1}\ell_J$) of physical   
$d$+$d$, $t$+$p$, and $h$+$n$ channels for $J \le 2$ and $\ell \le 2$. }
\begin{center}
\begin{tabular}{c|cccccc}
\hline
\backslashbox{channel}{$J^\pi$}    & 0$^+$ &1$^+$  &2$^+$  &0$^-$  &1$^-$  &2$^-$\\
\hline
$d(1^+)$+$d(1^+)$  &$^1S_0$&$^5D_1$&$^5S_2$&$^3P_0$&$^3P_1$&$^3P_2$\\
                   &$^5D_0$&       &$^1D_2$&&&\\
                   &       &       &$^5D_2$&&&\\
\hline
$t(\frac{1}{2}^+)$+$p(\frac{1}{2}^+), \;h(\frac{1}{2}^+)$+$n(\frac{1}{2}^+)$     &$^1S_0$&$^3S_1$&$^1D_2$&$^3P_0$&$^1P_1$&$^3P_2$\\
&       &$^3D_1$&$^3D_2$&       &$^3P_1$&       \\
\hline
\end{tabular}
\end{center}
\label{chan0}
\end{table}

Because one of our 
purposes in this investigation is to understand the role of the 
tensor force played in the four-nucleon dynamics, we want to 
compare the phase shifts obtained with two Hamiltonians that differ 
in the type of $NN$ interactions. 
One is a realistic interaction called 
the AV8$^{\prime}$ potential~\cite{av8p} that includes 
central, tensor and spin-orbit components. 
We also add an effective three-nucleon force (TNF) in order
to reproduce reasonably the binding energies of $t$, $h$ and $^4$He \cite{hiyama04}, which makes reasonable thresholds. In the present calculation,
the TNF is included in all calculations for AV8$^{\prime}$.
Another is an effective central interaction called
the Minnesota (MN) potential~\cite{thompson77},
which reproduce reasonably the binding energies of $t$, $h$ and $^4$He,
though it has 
central terms alone (with an exchange parameter $u=1$).
The Coulomb potential is included for both potentials. 

The intrinsic wave function $\Phi^{k}_{I_k}$ of cluster $k$ $(k=a,\ b)$ 
is described with a combination of $N_k$ basis functions with 
different $L_k$ and $S_k$ values  
\begin{eqnarray}
\Phi_{I_{k}M_{I_k}}^k=\sum^{N_k} {\cal A} \left[
\psi_{L_k}^{(\rm{space})}\psi_{S_k}^{(\rm{spin})} 
\right]_{I_k M_{I_k}}
\psi^{(\rm{isospin})}_{T_k M_{T_k}},
\label{wf.cluster}
\end{eqnarray}
where  $\psi_{L_k}^{(\rm{space})}$,
$\psi_{S_k}^{(\rm{spin})}$ and $\psi_{T_k M_{T_k} }^{(\rm{isospin})}$ 
denote the space,  spin and isospin parts of the cluster wave function.
In the case of the AV8$^{\prime}$ potential, the $t$ (or $h$) wave function 
is approximated with thirty Gaussian basis functions that include 
$L_k \leq 2$, and  $S_k=\frac{1}{2}$ and $\frac{3}{2}$. 
The deuteron wave function is also approximated with Gaussian basis
functions, four terms both in the $S$- and $D$-waves, respectively. 
The falloff parameters of the Gaussian functions are selected 
using the SVM~\cite{book} and the expansion coefficients are 
determined by diagonalizing the intrinsic cluster Hamiltonian. 
A similar procedure is applied to the case of the MN potential.

The calculated energies $E$, root-mean-square (rms) radii $R^{\rm rms}$ and
$D$ state probabilities $P_D$ 
are given in the fourth to sixth columns in Table~\ref{sub1}.
We use the truncated basis in order to obtain the phase shifts
in reasonable computer times,
they slightly deviate from more elaborate calculations,
which are given in the last three columns.
Fortunately, except for the small shift of the
threshold energy, the phase shifts are not very sensitive to the
details of the cluster wave functions because they are determined
by the change of the relative motion function of the clusters.
The $N_k$ values in parenthesis for $^4$He are the number of
$J^{\pi}=0^+$ basis functions in the major multi-channel calculation. 
The energy of $^4$He calculated in Table 2 with the multi-channel calculation  
is thus not optimized but found to be very close to that of
the more extensive calculation. 
It is noted that the calculated $R^{\rm rms}$ value for the deuteron is
smaller than in other calculations. This is due to the restricted
choice of the length parameters of the basis functions, which permits
us to use a relatively small channel radius of $a \sim 15$ fm. We have
checked that the phase shifts for a $d+d$ single channel
calculation do not change even when more extended deuteron wave functions 
are employed. 

\begin{table}
\caption{Energies $E$, rms radii $R^{\rm rms}$ and $D$-state probabilities $P_D$ 
of the clusters that appear in four-nucleon scattering and $^4$He
with the AV8$^{\prime}$ (with TNF) and MN potentials. 
$N_k$ is the number of basis functions used to approximate the wave function 
of cluster $k$. The values in the last three columns for
three- and four-body systems are taken 
from Ref.~\cite{hiyama04} for AV8$^{\prime}$ and Ref.~\cite{DGVR} for MN.}
\begin{center}
\begin{tabular}{ccccccccc}
\hline
potential&cluster& \multicolumn{4}{c}{present} & \multicolumn{3}{c}{literature} \\
& & $N_k$ &$E$&$R^{\rm rms}$&$P_D$& $E$&$R^{\rm rms}$&$P_{D}$\\ 
&& &(MeV)&(fm)&($\%$)&
(MeV)&(fm)&($\%$)\\ 
\hline
&$d(1^+)$  & 8    & $-$2.18 & 1.79  &5.9& $-$2.24 &1.96 &5.8\\
AV8$^{\prime}$&$t(\frac{1}{2}^+)$  & 30   & $-$8.22 &1.69  &8.4& $-$8.41 &- &-\\
(with TNF) &$h(\frac{1}{2}^+)$ & 30   & $-$7.55 & 1.71 &8.3& $-$7.74 &- &-\\
&$^4$He$(0^+)$ & (2370)   & $-$27.99 & 1.46  &13.8& $-$28.44 &- &14.1\\
\hline
&$d(1^+)$   & 4   & $-$2.10 & 1.63&0  & $-$2.20 & 1.95&0\\
MN&$t(\frac{1}{2}^+)$ & 15   & $-$8.38 & 1.70&0  & $-$8.38 & 1.71&0\\
&$h(\frac{1}{2}^+)$ & 15  & $-$7.70 & 1.72&0 & $-$7.71 & 1.74&0\\
&$^4$He$(0^+)$ & (1140)   & $-$29.94 & 1.41 &0& $-$29.94 & 1.41 &0\\
\hline
\end{tabular}
\end{center}
\label{sub1}
\end{table}

\section{Correlated Gaussian function with triple global vectors}
\label{sect.3}

As explained in the previous section, the calculation of the $S$-matrix 
reduces to that of the Hamiltonian and overlap matrix elements with the 
functions defined by (\ref{wf-ch1}) and (\ref{wf.cluster}), and it 
is conveniently performed by transforming 
that wave function into an $LS$-coupled form,
\begin{eqnarray}
&&{\cal A}
\left[\left[\left[\psi^{(\rm space)}_{L_a}\psi^{(\rm space)}_{L_b}\right]_{L_{ab}}
\chi_\alpha({\vi \rho}_\alpha)\right]_L
\left[\psi^{(\rm spin)}_{S_a} \psi^{(\rm spin)}_{S_b}\right]_{S}
\right]_{JM}.
\end{eqnarray}
The transformation can be done as
\begin{eqnarray}
& &{\hspace{-10mm}}
{\cal A}\left[ \left[
[\psi_{L_a}^{(\rm{space})}\psi_{S_a}^{(\rm{spin})}]_{I_a}
[\psi_{L_b}^{(\rm{space})}\psi_{S_b}^{(\rm{spin})}]_{I_b}
\right]_I \chi_\alpha({\vi \rho}_\alpha)\right]_{JM}
\psi^{(\rm{isospin})}_{T_a M_{T_a}}\psi^{(\rm{isospin})}_{T_b M_{T_b}}
\nonumber \\
&=&\sum_{L_{ab}LS}
\left[
\begin{array}{ccc}
L_a&S_a&I_a\\
L_b&S_b&I_b\\
L_{ab}&S&I\\
\end{array}\right]
(-1)^{L_{ab}+J-I-L}U(S L_{ab} J \ell; IL)\nonumber\\
&&\times
{\cal A}\left[
\psi_{L_a L_b (L_{ab}) \ell L}^{(\rm space)}
\left[\psi^{(\rm spin)}_{S_a} \psi^{(\rm spin)}_{S_b}\right]_{S}
\right]_{JM}
\psi^{(\rm{isospin})}_{T_a M_{T_a}}\psi^{(\rm{isospin})}_{T_b M_{T_b}}
\end{eqnarray}
with
\begin{eqnarray}
\psi_{L_a L_b (L_{ab}) \ell L}^{(\rm space)}
=\left[\left[\psi^{(\rm space)}_{L_a}\psi^{(\rm space)}_{L_b}\right]_{L_{ab}}
\chi_\alpha({\vi \rho}_\alpha)\right]_L,
\label{space.part}
\end{eqnarray}
where $U$  and 
$[ \ \ ]$ are Racah and 9$j$ coefficients in unitary form~\cite{book}. 

\begin{figure}[t]
\begin{center}
\epsfig{file=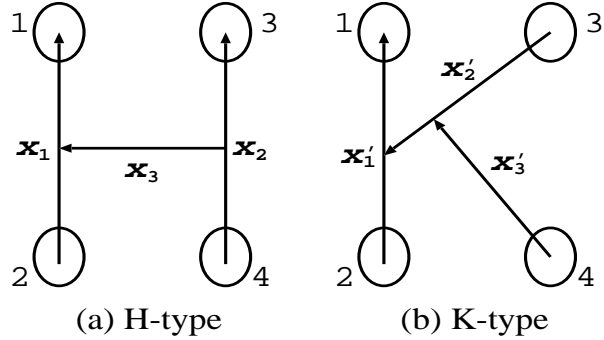,width=8.0cm,height=4.5cm}
\caption{Relative coordinates for the four-body system}
\label{fig.1}
\end{center}
\end{figure}

The evaluation of the matrix element can be done in the spatial, spin, 
and isospin parts separately. The spin and isospin parts are obtained 
straightforwardly. In the following 
we concentrate on the spatial matrix element. 
The spatial part~(\ref{space.part}) of the total wave function is given
as a product of the cluster intrinsic parts 
and their relative motion part. The coordinates used to describe the $2N$+$2N$
channel are depicted in Fig.~\ref{fig.1}(a) with ${\vi \rho}_\alpha={\vi x}_3$, 
whereas the coordinates suitable for the $t$+$p$ and $h$+$n$ channels are 
shown in Fig.~\ref{fig.1}(b) with ${\vi \rho}_\alpha={\vi x}_3'$. These
coordinate sets are often called H-type and K-type.  
Therefore the calculation of the spatial matrix element  requires 
a coordinate transformation involving the angular momenta $L_a, L_b,
\ell$ and $L_{ab}$. Moreover the permutation operator in ${\cal A}$
causes a complicated coordinate transformation. All these complexities 
are treated elegantly by introducing a correlated Gaussian~\cite{boys60,vs95,book}, provided 
each part of  $\psi_{L_a L_b (L_{ab}) \ell L}^{(\rm space)}$ is given 
in terms of (a combination of) Gaussian functions as in the present
case. In what follows we will demonstrate how it is performed. 
Because the formulation with the correlated Gaussian is not restricted
to four nucleons but can be applied to a system including more
particles, the number of nucleons is assumed to be 
$N$ in this and next sections as well as in Appendices B and C  
unless otherwise mentioned. 

The relative and center of mass coordinates of the $N$ nucleons, 
${\vi x}_i\, (i=1,\ldots,N)$, 
and the single-particle coordinates, ${\vi r}_i\, (i=1,\ldots,N)$, 
are mutually related by a linear transformation matrix $U$ and its 
inverse $U^{-1}$ as follows:
\begin{equation}
{\vi x}_i=\sum_{j=1}^{N}U_{ij}{\vi r}_j,\ \ \ \ \ 
{\vi r}_i=\sum_{j=1}^{N}(U^{-1})_{ij}{\vi x}_j.
\label{def.matU}
\end{equation}
We use a matrix notation as much as possible in order to simplify formulas and 
expressions. 
Let ${\vi x}$ denote an $(N\!-\!1)$-dimensional column vector comprising 
all ${\vi x}_i$ but the center of mass coordinate ${\vi x}_N$. 
Its transpose is a row vector and it is expressed as
\begin{equation}
\widetilde{{\vi x}}=({\vi x}_1,{\vi x}_2,...,{\vi x}_{N-1}).
\end{equation}
The choice for ${\vi x}$ is not unique but a set of Jacobi coordinates is 
conveniently employed.
For the four-body system, the Jacobi set is identical to the K-type
coordinate, and the 
corresponding matrix $U$ is given by 
\begin{eqnarray}
U_K=
\left(\begin{array}{rrrr}
1&-1&0&0\\
\frac{1}{2}&\frac{1}{2}&-1&0\\
\frac{1}{3}&\frac{1}{3}&\frac{1}{3}&-1\\
\frac{1}{4}&\frac{1}{4}&\frac{1}{4}&\frac{1}{4}\\
\end{array}\right).
\end{eqnarray}
The transformation matrix for  the H-type coordinate reads 
\begin{eqnarray}
U_H=
\left(\begin{array}{rrrr}
1&-1&0&0\\
0&0&1&-1\\
\frac{1}{2}&\frac{1}{2}&-\frac{1}{2}&-\frac{1}{2}\\
\frac{1}{4}&\frac{1}{4}&\frac{1}{4}&\frac{1}{4}\\
\end{array}\right).
\end{eqnarray}
The K-type coordinate is obtained directly from the 
H-type coordinate by a transformation matrix 
$U_K U_H^{-1}$
\begin{equation}
U_KU_H^{-1}=
\left(
\begin{array}{rrrr}
1&0&0&0\\
0&-\frac{1}{2}&1&0\\
0&\frac{2}{3}&\frac{2}{3}&0\\
0&0&0&1\\
\end{array}
\right)
=\left(
\begin{array}{cc}
U_{KH}&0\\
0& 1\\
\end{array}
\right),
\label{HtoK}
\end{equation}
where $U_{KH}$ is a 3$\times$3 sub-matrix of $U_K U_H^{-1}$.

Each coordinate set emphasizes particular correlations among the
nucleons. As mentioned above, 
the H-coordinate is natural to describe the $d$+$d$ channel, 
whereas the K-coordinate is suited for a description of the 3$N$+$N$ 
partition. It is 
of crucial importance to include both types of motion in order to 
fully describe the four-nucleon 
dynamics~\cite{benchmark}. In order to develop a unified method 
that can incorporate both 
types of coordinates on an equal footing, 
we extend the explicitly correlated Gaussian 
function~\cite{suzuki00,DGVR} to include triple 
global vectors
\begin{eqnarray}
&&F_{L_1 L_2 (L_{12})L_3  L M}(u_1,u_2,u_3,A,{\vi x})\nonumber\\
&&\quad  ={\rm exp}\left(-{\frac{1}{2}}{\widetilde{{\vi x}}} A {\vi x}\right)
\left[[{\cal Y}_{L_1}(\widetilde{u_1}{\vi x})
 {\cal Y}_{L_2}(\widetilde{u_2}{\vi x})]_{L_{12}}
{\cal Y}_{L_3}(\widetilde{u_3}{\vi x})\right]_{LM},
\label{cgtgv}
\end{eqnarray}
where 
\begin{equation}
{\cal Y}_{L_iM_i}(\widetilde{u_i}{\vi x})=
|\widetilde{u_i}{\vi x}|^{L_i}Y_{L_iM_i}(\widehat{\widetilde{u_i}{\vi x}})
\end{equation}
is a solid spherical harmonics and its argument, 
$\widetilde{u_i}{\vi x}$, what we call a global vector, is 
a vector defined through 
an $(N-1)$-dimensional column vector
$u_i$ and ${\vi x}$ as 
\begin{equation}
\widetilde{u_i}{\vi x}=\sum_{j=1}^{N-1}(u_i)_j{\vi x}_j,
\label{def.ux}
\end{equation}
where $(u_i)_j$ is the $j$th element of $u_i$. 
In Eq.~(\ref{cgtgv}) $A$ is  an $(N-1)\times(N-1)$ real 
and symmetric matrix, and it must be positive-definite for the 
function $F$ to have a finite norm, but 
otherwise may be arbitrary. Non-diagonal elements of $A$ can be nonzero.

The matrix $A$ and the vectors  
$u_1, u_2, u_3$ are parameters to characterize the ``shape'' 
of the correlated  Gaussian function.  The Gaussian function including
$A$ describes a spherical motion of the system, while the global vectors 
are responsible for a rotational motion. The spatial 
function~(\ref{space.part}) is found to reduce to the  
general form~(\ref{cgtgv}).  
Suppose that ${\vi x}$ stands for the H-type
coordinate. Then a choice of 
$\widetilde{u_1}$=(1,0,0), $\widetilde{u_2}$=(0,1,0)
and $\widetilde{u_3}$=(0,0,1) together with a diagonal matrix $A$ 
provides us with the 
basis function~(\ref{space.part}) employed to represent 
the configurations of the $2N$+$2N$ channel. 
On the other hand, the K-type basis function looks like 
\begin{equation}
{\rm exp}\left(-{\frac{1}{2}}{\widetilde{{\vi x}'}} A_K {\vi x}'\right)
\left[[{\cal Y}_{L_1}({\vi x}'_1)
 {\cal Y}_{L_2}({\vi x}'_2)]_{L_{12}}
{\cal Y}_{L_3}({\vi x}'_3)\right]_{LM},
\label{K.basis}
\end{equation}
where $\widetilde{{\vi x}'}=({\vi x}'_1,{\vi x}'_2,{\vi x}'_3)$ is the 
K-coordinate set (see Fig.~1(b)) and $A_K$ is a 3$\times$3 diagonal 
matrix. Noting that ${\vi x}'$ is equal to ${\vi x}'=U_{KH}{\vi x}$,  
we observe that 
the basis function~(\ref{K.basis}) is obtained 
from Eq.~(\ref{cgtgv}) by a particular choice of parameters, that is, 
$\widetilde{u_1}$=(1,0,0), $\widetilde{u_2}$=(0,$-\frac{1}{2}$,$1$)
and $\widetilde{u_3}$=(0,$\frac{2}{3}$,$\frac{2}{3}$), and 
the matrix $A$ is related to $A_K$ by 
\begin{eqnarray}
A=(u_1 u_2 u_3)A_K \left(
\begin{array}{c}
\widetilde{u_1}\\
\widetilde{u_2}\\
\widetilde{u_3}\\
\end{array}
\right)
=\widetilde{U_{KH}}A_K U_{KH}.
\end{eqnarray}
Thus the form of the $F$-function remains unchanged under the 
transformation of relative coordinates.

Note that $A$ is no longer diagonal. 
The choice of a different set 
of coordinates ends up only choosing appropriate parameters for $A$, $u_1$, $u_2$, and $u_3$. 

It is also noted that the triple global vectors in Eq.~(\ref{cgtgv}) 
are a minimum number of vectors to provide all possible spatial 
functions with arbitrary $L$ and parity $\pi$. 
A natural parity state with $\pi=(-1)^L$ can be described by only one 
global vector, that is, using e.g., $L_1=L$, $L_2=0$, $L_{12}=L$, 
$L_3=0$~\cite{vs95,suzuki98,GVR}. To
describe an unnatural parity state with $\pi=(-1)^{L+1}$ except for 
$0^-$ case, we need  at least two global vectors, say, 
$L_1=L$, $L_2=1$, $L_{12}=L$, $L_3=0$~\cite{suzuki00,DGVR}.  
The simplest choice for the $0^-$ 
state is  to use three global vectors with $L_1=L_2=L_{12}=L_3=1$~\cite{suzuki00}. 
In this way, the basis function~(\ref{cgtgv}) can be 
versatile enough to describe bound states of not only four- but also 
more-particle systems with arbitrary $L$ and $\pi$. 

To assure the permutation symmetry of the wave function, we have to operate a permutation $P$ on $F$. Since $P$ induces a linear 
transformation of the coordinate set, a new set of the permuted 
coordinates, ${\vi x}_P$, is related to the original 
coordinate set ${\vi x}$ as  
${\vi x}_P\!=\!{\cal P}{\vi x}$ with an $(N-1)\times (N-1)$ matrix ${\cal P}$. 
As before, this permutation does not change the form of the $F$-function:
\begin{eqnarray}
& &PF_{L_1 L_2(L_{12}) L_3 LM}(u_1,u_2,u_3,A,{\vi x})
\nonumber \\
&&\quad =F_{L_1 L_2 (L_{12})L_3 LM}(u_1,u_2,u_3,A,{\vi x}_P)
\nonumber \\ 
&&\quad =F_{L_1 L_2(L_{12}) L_3 LM}(\widetilde{\cal P}u_1,
\widetilde{\cal P}u_2,\widetilde{\cal P}u_3,\widetilde{\cal P}A{\cal P},
{\vi x}).
\label{trans.cg}
\end{eqnarray}
The fact that  
the functional form of $F$ remains unchanged under the permutation
as well as the transformation of coordinates
enables one to unify the method of calculating the matrix elements. 
This unique property is one of the most notable points in the present method.

\section{Calculation of matrix elements}
\label{sect.4}

Calculations of matrix elements with the correlated Gaussian $F$ are greatly facilitated with the aid of the generating function $g$~\cite{vs95,book}
\begin{eqnarray}
g({\vi s}; A, {\vi x})=
\exp\Big(-{\frac{1}{2}}{\widetilde{\vi x}}A{\vi x}+
\widetilde{\vi s}{\vi x}\Big),
\end{eqnarray}
with 
$\widetilde{\vi s}=({\vi s}_1,{\vi s}_2,\ldots,{\vi s}_{N-1})$, where  
${\vi s}_i=\sum_{j=1}^3\lambda_j(u_j)_i{\vi e}_j$,
${\vi e}_j$ is a 3-dimensional 
unit vector (${\vi e}_j\cdot{\vi e}_j=1$), and 
$\lambda_j$ is a scalar parameter. More explicitly 
\begin{equation}
\widetilde{\vi s}{\vi x}=\sum_{i=1}^{N-1}{\vi s}_i\cdot{\vi x}_i=
\sum_{i=1}^{N-1}\sum_{j=1}^3\lambda_j
(u_{j})_i {\vi e}_j \cdot{\vi x}_i
=\sum_{j=1}^3 \lambda_j{\vi e}_j\cdot (\widetilde{u_j}{\vi x}).
\end{equation}
The correlated Gaussian $F$ is generated as follows:
\begin{eqnarray}
&&F_{L_1 L_2(L_{12}) L_3 LM}(u_1,u_2,u_3,A,{\vi x})\nonumber\\
&&\quad =\left(\prod_{i=1}^3 \frac{B_{L_i}}{L_i !}\int d{\vi e}_i\right)
\left[\left[Y_{L_1}({\vi e}_1)Y_{L_2}({\vi e}_2)\right]_{L_{12}}
Y_{L_3}({\vi e}_3)\right]_{LM}\nonumber\\
&&\quad \times \left(\frac{\partial^{L_1+L_2+L_3}}
{\partial \lambda_1^{L_1} \partial \lambda_2^{L_2} \partial \lambda_3^{L_3} }
\,g({\vi s};A,{\vi x})\right)\Bigg\vert_{\lambda_1=\lambda_2=\lambda_3=0},\label{gfn}
\end{eqnarray}
where
\begin{eqnarray}
\hspace*{-2cm} B_L&=&\frac{(2L+1)!!}{4\pi}.
\end{eqnarray}
When $g({\vi s}; A,{\vi x})$ 
is expanded in powers of $\lambda_1$, only the term of degree 
$\lambda_{1}^{L_1}$ contributes in Eq.~(\ref{gfn}), and 
this term contains the $L_1$th degree ${\vi e}_1$ 
because $\lambda_1$ and 
${\vi e}_1$ always appear simultaneously. 
In order for the term 
to contribute to the integration over ${\vi e}_1$, these $L_1$ vectors 
${\vi e}_1$ must couple to the angular momentum $L_1$ because of the 
orthonormality of the spherical harmonics $Y_{L_1M_1}({\vi e}_1)$, that is, they 
are uniquely coupled to the maximum possible angular momentum. 
The same rule applies to $\lambda_2$, ${\vi e}_2$ and $\lambda_3$, ${\vi e}_3$  as well.

We outline a method of calculating 
the matrix element for an operator ${\cal O}$
\begin{equation}
\langle F_{L_4 L_5(L_{45}) L_6 L'M'}(u_4, u_5, u_6, A',{\vi x})\vert
 {\cal O}\vert F_{L_1 L_2(L_{12}) L_3 LM}(u_1, u_2, u_3, A,{\vi x})\rangle. 
\label{meofop}
\end{equation}
In what follows this matrix element is abbreviated as 
$\langle F'\vert {\cal O}\vert F\rangle $.  
Using Eq.~(\ref{gfn}) in Eq.~(\ref{meofop}) enables one to relate 
the matrix element to that between the generating functions: 
\begin{eqnarray}
\left<F'\vert {\cal O}  \vert F \right> 
&\!=\!&\left(\prod_{i=1}^6 \frac{B_{L_i}}{L_i !}\int d{\vi e}_i\right)
\left[\left[Y_{L_4}({\vi e}_4)Y_{L_5}({\vi e}_5)\right]_{L_{45}}
Y_{L_6}({\vi e}_6)\right]_{L'M'}^* \nonumber\\
&\!\times\!&\left[\left[Y_{L_1}({\vi e}_1)Y_{L_2}({\vi e}_2)\right]_{L_{12}}
Y_{L_3}({\vi e}_3)\right]_{LM}
\nonumber\\
&\!\times\!&\left(\prod_{i=1}^6
\frac{\partial^{L_i}}{\partial \lambda_i^{L_i}}\right)
\left<g({\vi s}',A',{\vi x}')\vert {\cal O}  \vert
g({\vi s},A,{\vi x})\right>\Big\vert_{\lambda_i=0},
\label{grandformula}
\end{eqnarray}
with
\begin{equation}
{\vi s}={\lambda}_1u_1{\vi e}_1 \!+\!{\lambda}_2u_2{\vi e}_2
\!+\!{\lambda}_3u_3{\vi e}_3,\hspace*{1cm}
{\vi s}'={\lambda}_4u_4{\vi e}_4 \!+\!{\lambda}_5u_5{\vi e}_5
\!+\!{\lambda}_6u_6{\vi e}_6.
\end{equation}
The calculation of the matrix element consists of three stages: 
(1) Evaluate the matrix element between the generating functions, 
$\left<g({\vi s}',A',{\vi x}')\vert {\cal O}  \vert
g({\vi s},A,{\vi x})\right>$. (2) Expand that matrix element in 
powers of $\lambda_i$ and keep only those terms of degree $L_i$ for 
each $i$. (3) Recouple 
the vectors ${\vi e}_i$ and integrate over the angle coordinates. 
In the second stage the remaining terms should contain
${\vi e}_i$s of degree $L_i$ as well.  Hence  
any term with $\lambda_i^2 {\vi e}_i\!\cdot\!{\vi e}_i\!
=\!\lambda_i^2$ etc. can be omitted because the degree of ${\vi e}_i$ 
becomes smaller than that of $\lambda_i$.

We will explain the above procedures for the case of an overlap matrix 
element. The matrix element between the generating functions is 
\begin{eqnarray}
\left<g({\vi s}',A',{\vi x}')\vert 
g({\vi s},A,{\vi x})\right> 
=\left(\frac{(2\pi)^{N-1}}{\mbox{det} B}\right)^{3/2}
\exp\left(\frac{1}{2}\widetilde{{\vi z}}B^{-1}{\vi z}\right)
\end{eqnarray}
with
\begin{eqnarray}
B=A'+A,\qquad {\vi z}={\vi s}+{\vi s}'=\sum_{i=1}^6\lambda_i{\vi e}_iu_i.
\end{eqnarray}
To perform the operation in the second stage we note that 
\begin{equation}
\frac{1}{2}\widetilde{{\vi z}}B^{-1}{\vi z}=\frac{1}{2}\sum_{i,j=1}^6
\rho_{ij}\lambda_i\lambda_j {\vi e}_i\cdot{\vi e}_j
\end{equation}
with
\begin{equation}
\rho_{ij}=\widetilde{u_i}B^{-1}u_j.
\label{def.rho}
\end{equation}
As mentioned above, here we can drop the diagonal terms, $\lambda_i^2
{\vi e}_i\cdot{\vi e}_i$, and we get  
\begin{eqnarray}
&&\left(\prod_{i=1}^6\frac{\partial^{L_i}}{\partial \lambda_i^{L_i}}\right)
\left<g({\vi s}',A',{\vi x}')\vert 
g({\vi s},A,{\vi x})\right>\Big\vert_{\lambda_i=0} \nonumber\\
&&\quad =\left(\frac{(2\pi)^{N-1}}{\mbox{det} B}\right)^{3/2}\prod_{i=1}^6L_i!
\prod_{i<j}^6\frac{\left(\rho_{ij}{\vi e}_i\cdot{\vi e}_j\right)^{n_{ij}}}{n_{ij}!}.
\end{eqnarray}
Here the non-negative integers $n_{ij}$ must satisfy the 
following equations 
in order to assure the degree $L_i$ for ${\vi e}_i$ in the different terms,
\begin{eqnarray}
&&n_{12}+n_{13}+n_{14}+n_{15}+n_{16}=L_1, \nonumber\\
&&n_{12}+n_{23}+n_{24}+n_{25}+n_{26}=L_2, \nonumber\\
&&n_{13}+n_{23}+n_{34}+n_{35}+n_{36}=L_3, \nonumber\\
&&n_{14}+n_{24}+n_{34}+n_{45}+n_{46}=L_4, \nonumber\\
&&n_{15}+n_{25}+n_{35}+n_{45}+n_{56}=L_5, \nonumber\\
&&n_{16}+n_{26}+n_{36}+n_{46}+n_{56}=L_6. 
\label{nij}
\end{eqnarray}

The last step is to recouple the angular momenta arising 
from the various terms. Since we have to couple ${\vi e}_i$s to the 
angular momentum $L_i$ from the terms of degree $L_i$, we may 
replace the term $\left(\rho_{ij}{\vi e}_i\cdot{\vi e}_j\right)^{n_{ij}}$ 
with just one piece 
\begin{equation}
\frac{ (-\rho_{ij})^{n_{ij}} n_{ij}! \sqrt{2n_{ij}+1}}{B_{n_{ij}}}
\left[Y_{n_{ij}}({\vi e}_i)Y_{n_{ij}}({\vi e}_j)\right]_{00}.
\end{equation}
Other pieces like $\left[Y_{\kappa}({\vi e}_i)Y_{\kappa}({\vi
e}_j)\right]_{00}$ with $\kappa < n_{ij}$ do not contribute to the 
matrix element. 
We thus have a product of 15 terms of $\left[Y_{n_{ij}}({\vi e}_i)Y_{n_{ij}}({\vi e}_j)\right]_{00}$. The coupling of these terms is done by 
defining various coefficients that are all expressed 
in terms of Clebsch-Gordan, Racah, and 9$j$ coefficients. 
For example, we make use of the formulas 
\begin{eqnarray}
& &{\hspace{-1cm}}[Y_a({\vi e}_1)Y_a({\vi e}_2)]_{00}\ [Y_b({\vi e}_1)Y_b({\vi e}_3)]_{00}\ 
[Y_c({\vi e}_2)Y_c({\vi e}_3)]_{00}
\nonumber \\
& \to& X(abc)\ [[Y_{a+b}({\vi e}_1)Y_{a+c}({\vi e}_2)]_{b+c}Y_{b+c}({\vi e}_3)]_{00},\\
\nonumber\\
& &{\hspace{-1cm}}[Y_a({\vi e}_1)Y_a({\vi e}_4)]_{00}\ [Y_b({\vi e}_1)Y_b({\vi e}_5)]_{00}\ 
[Y_c({\vi e}_1)Y_c({\vi e}_6)]_{00}
\nonumber \\
&\to& R_3(abc)\ [Y_{a+b+c}({\vi e}_1)\ [[Y_{a}({\vi e}_4)Y_{b}({\vi e}_5)]_{a+b}Y_c({\vi e}_6)]_{a+b+c}]_{00}.
\end{eqnarray}
Here the symbol $\to$ indicates that no other terms arising from the 
left hand side of the equation contribute to the integration over the 
angles ${\vi e}_i$s, so that only the term on the right hand side has 
to be retained. Another coefficient is 
\begin{eqnarray}
&&{\hspace{-1cm}}[[[Y_a({\vi e}_4)Y_b({\vi e}_5)]_qY_c({\vi e}_6)]_Q\ 
[[Y_{a'}({\vi e}_4)Y_{b'}({\vi e}_5)]_{q'}Y_{c'}({\vi e}_6)]_{Q'}]_{\ell}
\nonumber \\
&\to& \sum_{\ell'}W(abcqQ,a'b'c'q'Q',\ell \ell')
[[Y_{a+a'}({\vi e}_4)Y_{b+b'}({\vi e}_5)]_{\ell'}Y_{c+c'}({\vi e}_6)]_{\ell}.
\end{eqnarray}
Expressions for the coefficients, $X, R_3, W$, are given in Appendix A. 
Performing the integration of the six unit vectors, ${\vi e}_i$s,
as prescribed in Eq.~(\ref{grandformula}) leads to the overlap matrix 
element
\begin{eqnarray}
& &{\hspace{-1cm}}\left<F'\vert  F \right>
\nonumber \\
&=&\left(\frac{(2\pi)^{N-1}}{\mbox{det} B}\right)^{3/2}
\left(\prod_{i=1}^6 B_{L_i}\right)
\frac{(-1)^{L_1+L_2+L_3}}{\sqrt{2L+1}}\delta_{LL'}\delta_{MM'}\nonumber\\
&\times & \hspace*{-0.3cm}\sum_{n_{ij}}\left(\prod_{i<j}^6 (-\rho_{ij})^{n_{ij}}
\frac{ \sqrt{2n_{ij}+1}}{B_{n_{ij}}}\right)
O(n_{ij}; L_1L_2L_3L_4L_5L_6,L_{12}L_{45}L)\label{eq6a},
\label{me.overlap}
\end{eqnarray}
with
\begin{eqnarray}
&&{\hspace{-1cm}}O(n_{ij}; L_1L_2L_3L_4L_5L_6,L_{12}L_{45}L)\nonumber\\
&&{\hspace{-1cm}}=X(n_{12}n_{13}n_{23})R_3(n_{14}n_{15}n_{16})R_3(n_{24}n_{25}n_{26})
R_3(n_{34}n_{35}n_{36})X(n_{45}n_{46}n_{56})\nonumber\\
&&{\hspace{-1cm}}\times Z(n_{12}\!+\!n_{13}\ L_1\!-\!n_{12}\!-\!n_{13})
Z(n_{12}\!+\!n_{23}\ L_2\!-\!n_{12}\!-\!n_{23})
Z(n_{13}\!+\!n_{23}\ L_3\!-\!n_{13}\!-\!n_{23})\nonumber\\
&&{\hspace{-1cm}}\times \sum_{\ell_1 \ell_2 \ell_3}
\left[\begin{array}{ccc}
L_1&L_1\!-\!n_{12}\!-\!n_{13}&n_{12}\!+\!n_{13}\\
L_2&L_2\!-\!n_{12}\!-\!n_{23}&n_{12}\!+\!n_{23}\\
L_{12}&\ell_1&n_{13}\!+\!n_{23}\\
\end{array}\right]
\left[\begin{array}{ccc}
L_{12}&\ell_1&n_{13}\!+\!n_{23}\\
L_3&L_3\!-\!n_{13}\!-\!n_{23}&n_{13}\!+\!n_{23}\\
L&L&0\\
\end{array}\right]
\nonumber\\
&&{\hspace{-1cm}}\times W(n_{14}n_{15}n_{16}\ n_{14}\!+\!n_{15}\ L_1\!-\!n_{12}\!-\!n_{13},
n_{24}n_{25}n_{26}\ n_{24}\!+\!n_{25}\ L_2\!-\!n_{12}\!-\!n_{23}, \ell_1 \ell_2)\nonumber\\
&&{\hspace{-1cm}}\times W(n_{14}\!+\!n_{24}\ n_{15}\!+\!n_{25}\ n_{16}\!+\!n_{26}\ \ell_2\ \ell_1,
n_{34}n_{35}n_{36}\ n_{34}\!+\!n_{35}\ L_3\!-\!n_{13}\!-\!n_{23}, L\ell_3) \nonumber\\
&&{\hspace{-1cm}}\times W(L_4\!-\!n_{45}\!-\!n_{46}\ L_5\!-\!n_{45}\!-\!n_{56}\ L_6\!-\!n_{46}\!-\!n_{56}\ \ell_3\ L, \nonumber\\
&&\hspace*{3cm}
n_{45}\!+\!n_{46}\ n_{45}\!+\!n_{56}\ n_{46}\!+\!n_{56}\ n_{46}\!+\!n_{56}\ 0, L L_{45}),
\end{eqnarray}
where  $Z$ is 
the coefficient given in Eq.~(\ref{def.z}). 
The summation in Eq.~(\ref{me.overlap}) extends over all possible 
sets of $n_{ij}$ that satisfy 
Eq.~(\ref{nij}). In most cases the values of $L_i$ are limited up to 2, 
so that 
the number of terms to be evaluated is not so large and the 
calculation of the matrix element is fast. 

Expressions for the Hamiltonian matrix elements are collected in Appendix B. 
One advantage of our method is that the calculation of matrix elements can be 
done analytically. In addition we do not need to 
do angular momentum and parity projections because the correlated 
Gaussian function~(\ref{cgtgv}) already preserves those  
quantum numbers. 

The Fourier transform of the correlated Gaussian function $F$ is a
momentum space function and it becomes a 
useful tool to calculate various matrix elements that depend on the momentum 
operators~\cite{DGVR}. For example, the distribution of the relative momentum 
is obtained by the expectation value of $\delta({\vi p}_i-{\vi p}_j-{\vi
p})$, where ${\vi p}_j$ is the momentum of the $j$th particle. 
It is obviously much easier to calculate the distribution using the 
momentum space function rather than  the coordinate space function.
We show in Appendix C that  the Fourier transform of $F$ 
reduces to a linear combination of $F$s in the momentum space.

\begin{figure}[b]
\centerline{\includegraphics[width=9.0 cm,height=8.0 cm]{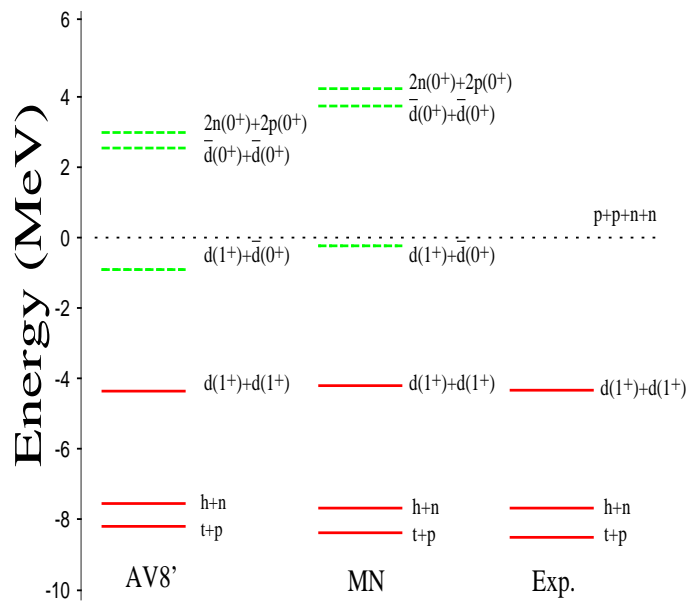}}
\caption{
Two-body thresholds calculated with the AV8$^{\prime}$ (left) and MN (middle)
potentials. The solid
lines are physical channels and the dashed lines are pseudo
channels. We also plot experimental two-body thresholds for physical channels (right).
The dotted line is the $p$+$p$+$n$+$n$ threshold.}
\label{fig:threshold}
\end{figure}

\section{Results}	
\label{results}
\subsection{$2N$+$2N$ and $3N$+$N$ channels}	
In Table \ref{chan0}, we gave the physical channels, $d$+$d$, $t$+$p$,
and $h$+$n$.
Fig.~\ref{fig:threshold} displays two-body decay
thresholds in the $d$+$d$ threshold energy region.
The three physical channels are the main channels that describe the scattering
around the three lowest thresholds ($d$+$d$, $t$+$p$, $h$+$n$).
However, the scattering wave function 
$\Psi^{JM\pi}_{\rm int}$ in the internal region should contain all effects that
may occur when all the nucleons come close to each other. It is thus
reasonable that $\Psi^{JM\pi}_{\rm int}$ may not be well described
in terms of the physical channels alone. Particularly the deuteron
can be easily distorted when we use realistic potentials. 

We will show that some pseudo $2N$+$2N$ channels
are indeed needed
to simulate the distortion of the deuteron.
These pseudo channels, when they are included in the phase-shift
calculation, are expected to take account of the distortion of
the clusters of the entrance channel \cite{kanada85}.
Here ``pseudo'' means that the clusters in the
pseudo channels are not physically observable but may play a
significant role in the internal region. The wave functions of
these $2N$ pseudo clusters are obtained by diagonalizing the intrinsic
cluster Hamiltonian similarly to the case of the physical clusters.
We take into account the following pseudo clusters: $d^*(1^+, T=0),
\ d^*(0^+, T=1), \ d^*(2^+, T=0), \ d^*(3^+, T=0)$, $2n^*(0^+, T=1)$, and   
$2p^*(0^+,T=1)$, where the upper suffix * indicates all the excited state but the
ground state of $d$. 
Among the pseudo clusters,
the lowest energy states with 0$^+$ that are related to
virtual states would be most important.
We especially write them as  $\bar{d}(0^+)$, 2$n(0^+)$ (di-neutron) and 2$p(0^+)$ (di-proton).
Although they are not bound, they are observed as resonances or
quasi-bound states with negative scattering lengths.
In fact the scattering lengths are 
$a_s(nn)=-$16.5 fm and $a_s(pp)=-$17.9 fm, which are comparable to 
$a_s(np, T=1)=-$23.7 fm.
The calculated thresholds of these pseudo channels are also drawn in
Fig.~\ref{fig:threshold}.

\begin{table}
\caption{$2N$+$2N$ and $3N$+$N$ channels. 
The Roman and Arabic numerals correspond to sets of channels included in the calculations. }
\begin{center}
\begin{tabular}{|c|c|c|c|}
\hline
\multicolumn{3}{|c|}{model}&channel\\
\hline
&$2N$+$2N$&  I&$d(1^+)$+$d(1^+)$\\
&&   &$d(1^+)$+$d^*(1^+)$\\
&&   &$d^*(1^+)$+$d^*(1^+)$\\
&& II&$\bar{d}(0^+)$+$\bar{d}(0^+)$\\
&&   &$\bar{d}(0^+)$+$d^*(0^+)$\\
&&   &$d^*(0^+)$+$d^*(0^+)$\\
&&III&$d^*(2^+)$+$d^*(1^+)$\\
&&   &$d^*(2^+)$+$d^*(2^+)$\\
&& IV&$d^*(3^+)$+$d^*(1^+)$\\
FULL&&   &$d^*(3^+)$+$d^*(2^+)$\\
&&   &$d^*(3^+)$+$d^*(3^+)$\\
&&  V&$2n(0^+)$+$2p(0^+)$\\
&&   &$2n(0^+)$+$2p^*(0^+)$\\
&&   &$2n^*(0^+)$+$2p(0^+)$\\
&&   &$2n^*(0^+)$+$2p^*(0^+)$\\
\cline{2-4}
&$3N$+$N$&1&$t(\frac{1}{2}^+)$+$p(\frac{1}{2}^+)$\\
&&   &$t^*(\frac{1}{2}^+)$+$p(\frac{1}{2}^+)$\\
&&  2&$h(\frac{1}{2}^+)$+$n(\frac{1}{2}^+)$\\
&&   &$h^*(\frac{1}{2}^+)$+$n(\frac{1}{2}^+)$\\
\hline
\end{tabular}
\end{center}
\label{chan1}
\end{table}

Though it is expected that the pseudo channels with 
low threshold energies contribute more strongly to the scattering phase
shift, we take into account all of these $2N$+$2N$ channels that include
a vanishing total isospin as given in Fig.~\ref{fig:threshold}.
The total isospin of the $3N$+$N$ channel is 
mixed in the present calculation. Because 
the  $T$=1 component of the scattering wave function only weakly 
couples to the $d(1^+, T=0)$+$d(1^+, T=0)$ elastic-channel,  
the channel $d(1^+, T=0)$+$\bar{d}(0^+, T=1)$ is
not employed in the calculation.

We also include the excited deuteron channels that comprise
the $d^*(2^+, T=0)$ and  $d^*(3^+, T=0)$ clusters.
The energies of these lowest thresholds
are above 10 MeV. 
These channels are therefore expected not to be very important, 
but that is not always the case as will be discussed 
in the case of the $^1$S$_0$ $d$+$d$
phase shift. 

Table~\ref{chan1} summarizes all the channels that are used in our 
calculation. The 2$N$+2$N$ channels are distinguished by 
Roman numerals, while the 3$N$+$N$ channels are labeled by 
Arabic numerals. In the following, we use an abbreviation 
``2$N$+2$N$'' or ``3$N$+$N$'' 
to indicate calculations including all 2$N$+2$N$ channels 
I-V or all 3$N$+$N$  channels (1-2 in Table~\ref{chan1}), respectively. 
Here $t^*(\frac{1}{2}^+)$ and $h^*(\frac{1}{2}^+)$ are excited
3$N$ continuum states.
A ``FULL'' calculation indicates that all the channels in the table are
included to set up the $S$-matrix. In  the case of the MN potential 
channels III and IV are not included because this potential contains no tensor 
force. 

The relative wave functions 
$\chi_{\alpha m}$ are expanded with 15 basis functions.
We checked the stability of the $S$-matrix against the choice of 
the channel radius. The channel radius employed 
in this calculation is about 15 fm. 

\subsection{Positive parity phase shifts}

\begin{figure}[b]
\centerline{\includegraphics[width=7.0 cm,height=6.0 cm]{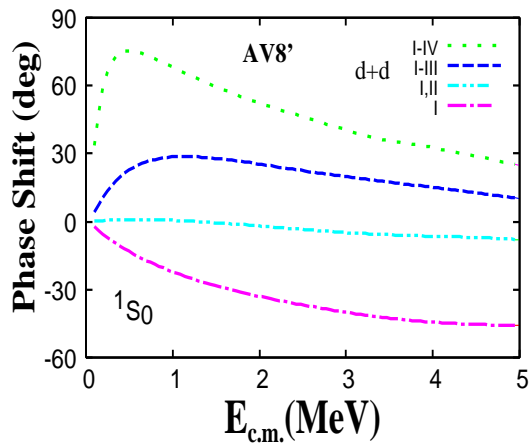}}
\caption{
$^1S_0$ $d$+$d$ elastic-scattering phase shift calculated
with the AV8$^{\prime}$ potential. The phase shifts are all obtained within the
$d$+$d$  channels. The set of included channels is successively increased from
I to IV. See Table~\ref{chan1} for the deuteron states included in
each channel.
} 
\label{fig:d+d-g3}
\end{figure}

Fig.~\ref{fig:d+d-g3} displays the $^1S_0$ $d$+$d$ elastic-scattering
phase shift obtained with the AV8$^{\prime}$ potential.
The dash-dotted line is the phase shift calculated with
channel I ($I_d=1^+$), and
the dash-dot-dotted line is the phase shift with channels I and II
($I_d=1^+, 0^+$). The phase shifts calculated by including further excited
deuterons are also plotted by the dashed and dotted lines that
correspond to the channels I-III ($I_d\le2^+$) and
I-IV ($I_d\le3^+$), respectively. A naive expectation that the
$^1S_0$ $d$+$d$ elastic-scattering phase shift might be well described in
channel I ($d(1^+)+d(1^+)$, $d(1^+)+d^*(1^+)$  and $d^*(1^+)+d^*(1^+)$) alone completely
breaks down in the case of the AV8$^{\prime}$ potential.  

Because the deuteron has a virtual state $\bar{d}$
with $0^+$ at low
excitation energy, it is reasonable that the inclusion of channel II
gives rise to a considerable attractive effect of several tens of
degrees on the phase shift, as shown by the dash-dot-dotted
line of Fig.~\ref{fig:d+d-g3}.
However, the phase shift exhibits no converging behavior even when
the higher spin states such as $d^*$(2$^+$) and $d^*$(3$^+$) are taken
into account in the calculation.
The additional attractions by these channels
are of the same order as that of channel II. One may conclude that
the deuteron is strongly distorted even in the low energy $^1S_0$
$d$+$d$ elastic scattering but more physically we have to realize
that there exist two observed $0^+$ states below the $d$+$d$
threshold. Obviously the $d$+$d$ scattering wave function is
subject to the structure of those states in the
internal region. 

The second 0$^+$ state of $^4$He lying about 4 MeV below
the $d$+$d$ threshold is known to have a $3N$+$N$ cluster
structure~\cite{horiuchi08,hiyama04}. Thus this state together with the
ground state of $^4$He cannot be described well in the $2N$+$2N$ model
space alone. As seen in Table~\ref{chan0}, the $3N$+$N$ channel contains a 
$^1S_0$ component, which is the dominant component of the $0^+_2$ state.
Since the realistic force strongly couples the $2N$+$2N$ channel
to the $3N$+$N$
channel and the $d$+$d$ scattering wave function has to be orthogonal to
the main component of the underlying $0^+$ states, we expect that
the deuteron in the incoming $d$+$d$ channel never remains in its ground
state but has to be distorted largely due to the $3N$+$N$
channel. The phase shift for 
the channel I-IV (dotted line) shows a resonant pattern.
This resonant state is expected to be the second 0$^+$ state
because of the restricted model space within the $d$+$d$ channel.

\begin{figure}[bh]
\centerline{\includegraphics[width=10.0 cm,height=8.0 cm]{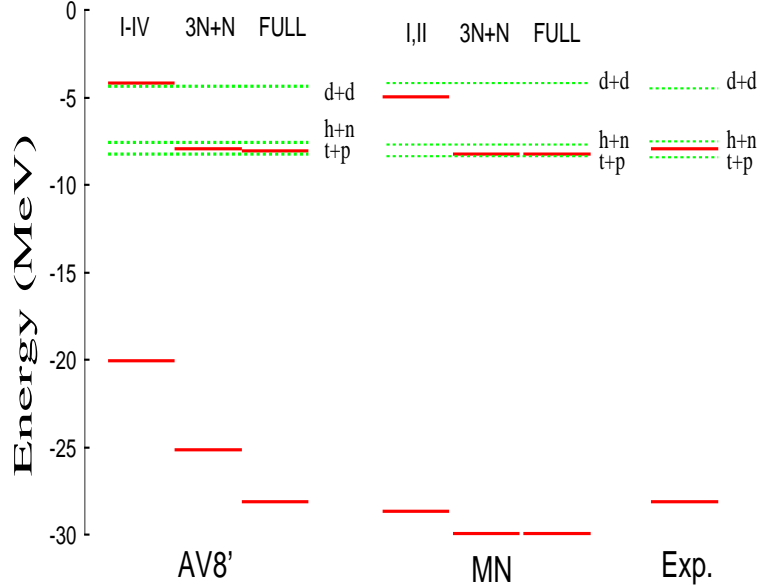}}
\caption{Comparison of the ground and second 0$^+$ state energies
between calculations with  
the AV8$^{\prime}$ (left) and MN (middle) potentials and experiment (right).
The model space for AV8$^{\prime}$ is I-IV, $3N$+$N$ and FULL
and the model space for MN is I-II, $3N$+$N$ and FULL.
}
\label{fig:2nd0+}
\end{figure}

\begin{figure}[ht]
\centerline{\includegraphics[width=6.0 cm,height=12.0 cm,angle=-90]{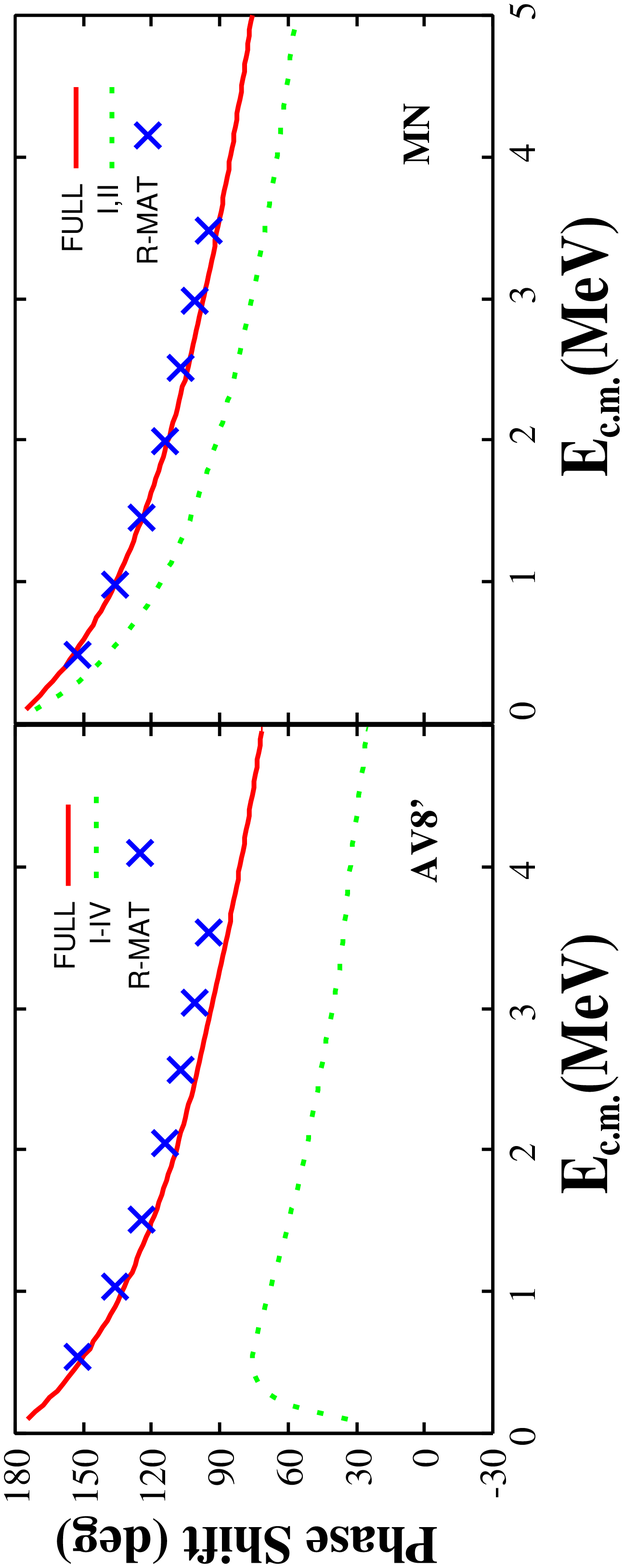}}
\caption{
$^1S_0$ $d$+$d$ elastic-scattering phase shift calculated
with the AV8$^{\prime}$ (left) and MN (right) potentials.
The solid line is a FULL calculation, while
the dotted line is a $d$+$d$ channel calculation. 
Crosses correspond to the $R$-matrix analysis of Ref.~\cite{hofmann08}.}
\label{fig:d+d-3N+N}
\end{figure}

Fig.~\ref{fig:2nd0+} displays the calculated ground state energy and the second 0$^+$
energy for the AV8$^{\prime}$ (left) and MN (middle) potentials.
The model spaces of the calculations
are I-IV, $3N$+$N$ and FULL for AV8$^{\prime}$ and
I-II, $3N$+$N$ and FULL for MN. We also plot experimental energies (right) \cite{tilley92}.
For the AV8$^{\prime}$ potential, the energies of the two lowest $0^+$ states 
do not change very much between the FULL and $3N$+$N$ models. But
the second 0$^+$ state with the $d$+$d$ model (channels I-IV)
is not bound with respect to the $d$+$d$ threshold as expected before. On the contrary,
for the MN potential,
the second 0$^+$ state with the $d$+$d$ model (channels I-II)
is bound with respect to the $d$+$d$ threshold.
We consider that this difference makes the drastic change of the  $d$+$d$ phase shifts,
between the AV8$^{\prime}$ and MN potentials.
It is also interesting to see that 
the energies of the two lowest $0^+$ states for the MN potential
are almost the same between the FULL and $3N$+$N$ models.

Plotted in Fig.~\ref{fig:d+d-3N+N} are
the $^1S_0$ $d$+$d$ elastic-scattering phase shifts obtained
with the AV8$^{\prime}$ potential (left) and the MN potential (right).
The FULL calculation (solid line) couples all 2$N$+2$N$ and 3$N$+$N$ channels that
are listed in Table~\ref{chan1}. 
The $R$-matrix analysis (crosses) \cite{hofmann08} is reproduced well by
both the AV8$^{\prime}$ and  MN potential with the FULL calculation.
Compared to the uncoupled phase shift
(dotted line), one clearly sees that the 3$N$+$N$ channel
produces a very large effect
on the $d$+$d$ elastic phase shift, especially in the case of the 
AV8$^{\prime}$ potential. We also verified that a calculation excluding
the channels III, IV or V from the FULL channel calculation gives
only negligible change in the phase shift.
The slow convergence seen in Fig.~\ref{fig:d+d-g3} is thus attributed
to the neglect of the $3N$+$N$ channel, indicating that
a proper account of the $^1S_0$ $d$+$d$ elastic phase shift at low
energy can be possible only when the coupled channels 
\{$d$(1$^+$)+$d$(1$^+$)\} +\{$d$(0$^+$)+$d$(0$^+$)\}+
\{$t$(1/2$^+$)+$p(1/2^+)$\} +\{$h$(1/2$^+$)+$n(1/2^+)$\} are
considered.

Thus, the slow convergence in Fig.~\ref{fig:d+d-g3} suggests that
the $2N$+$2N$ partition is not an economical way to include the effects
of the $3N$+$N$ channel.
In the case of the MN potential (right panel in Fig.~\ref{fig:d+d-3N+N}),
the situation is very different from the AV8$^{\prime}$ case.
The channel coupling effect is rather modest, and the size of the $^1S_0$
$d$+$d$ elastic phase shift is already accounted for mostly in the
$d$+$d$ channel calculation. 
All these results are very consistent
with the $0^+$ spectrum in Fig.~\ref{fig:2nd0+}.

\begin{figure}[ht]
\centerline{\includegraphics[width=6.0 cm,height=12.0 cm,angle=-90]{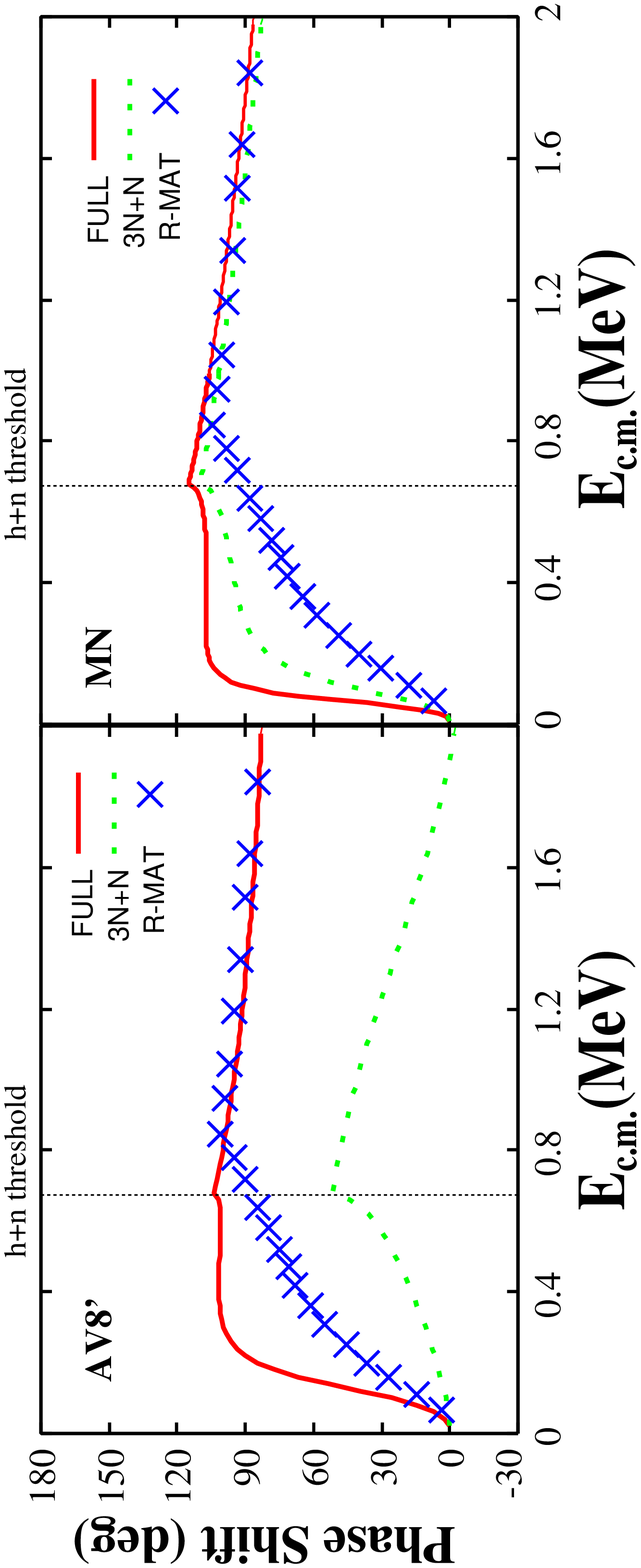}}
\caption{
$^1S_0$ $t$+$p$ elastic-scattering phase shift at 
energies below the
$d$+$d$ threshold. The solid line is a FULL calculation, while
the dashed line is a $3N$+$N$ channel calculation.
Crosses correspond to the $R$-matrix analysis of Ref.~\cite{hofmann08}.
}
\label{fig:3N+N-g3mn}
\end{figure}

The large distortion effect of the deuteron clusters
on the $^1S_0$ $d$+$d$ scattering phase shift
is expected to appear in the 3$N$+$N$ phase shift as well because of
the coupling between the 3$N$+$N$ and 2$N$+2$N$ channels.
We display in Fig.~\ref{fig:3N+N-g3mn}
the $^1S_0$ $t$+$p$ elastic-scattering
phase shift at energies below the $d$+$d$ threshold.
The 0$^+_2$ state of $^4$He is observed as a sharp resonance with a
proton decay width of 0.5 
MeV  at about 0.4 MeV above the $t$+$p$ threshold.
The present energies ($E_r=0.15$ MeV for AV8$^{\prime}$, $E_r=0.12$ MeV for MN) calculated
with a bound state approximation are 
slightly smaller than the experimental value, 
but they are consistent with a calculation ($E_r=0.105$ MeV and  $\Gamma/2=0.129$ MeV for AV18+UIX, $E_r=0.091$ MeV and  $\Gamma/2=0.077$ MeV for AV18+UIX+V$^*_3$)
with another realistic interaction (AV18) with three nucleon forces by Hofmann and Hale \cite{hofmann08}.
The calculated phase shifts 
appear slightly larger than that in the $R$-matrix analysis (crosses
in Fig.~\ref{fig:3N+N-g3mn}) \cite{hofmann08}.
It is noted that the phase shift changes so much even for a small
change of the 0$^+_2$ resonant pole position ($\sim$0.1 MeV)
because it is very near to the threshold.
The phase shifts in the FULL
calculation, for both AV8$^{\prime}$ and MN cases, show a resonance pattern in a
small energy interval and the overall energy dependencies of the phase shifts
are similar to each other.
However, the phase shifts
obtained only in the $3N$+$N$ channel are quite
different as indicated by the dotted lines in Fig.~\ref{fig:3N+N-g3mn}.
In the case of the MN potential (right) the
phase shift is already close to the FULL phase shift, while
in the case of the AV8$^{\prime}$ potential (left) the phase shift is much smaller
(by almost 90 degrees) and moreover shows no
resonance pattern. 

By looking into the wave functions in more detail,
we argue that the large distortion effect in the $^1S_0$ $d$+$d$
and $3N$+$N$ coupled channels is really brought about by the
tensor force. As shown in
Table ~\ref{sub1}, the AV8$^{\prime}$ potential with TNF gives 5.8\% and 8.4\% (8.3\%)
$D$-state probability for $d$ and $t$ $(h)$, respectively.
Thus the $d$+$d$ state in the $^1S_0$ state
contains $L=S=0$ components (89\%) as well as  $L=S=2$
components (11\%), where $L$ and $S$ are the total orbital and spin
angular momenta of the four-nucleon system. Similarly the $3N$+$N$
state in the $^1S_0$
state contains an $L=S=0$ component (92\%) and an $L=S=2$ component
(8\%). Thus the tensor force couples both states with $\Delta L=2$
and $\Delta S=2$ couplings, which are in fact very large compared to
the central matrix element ($\Delta L=0$, $\Delta S=0$). An analysis of
this type was performed for some levels of $^4$He in
Refs.~\cite{DGVR,horiuchi08}.
The MN potential contains no tensor force, so that the $d$+$d$ and
$3N$+$N$ channel coupling is modest.

\begin{figure}[t]
\centerline{\includegraphics[width=6.0 cm,height=13.0 cm,angle=-90]{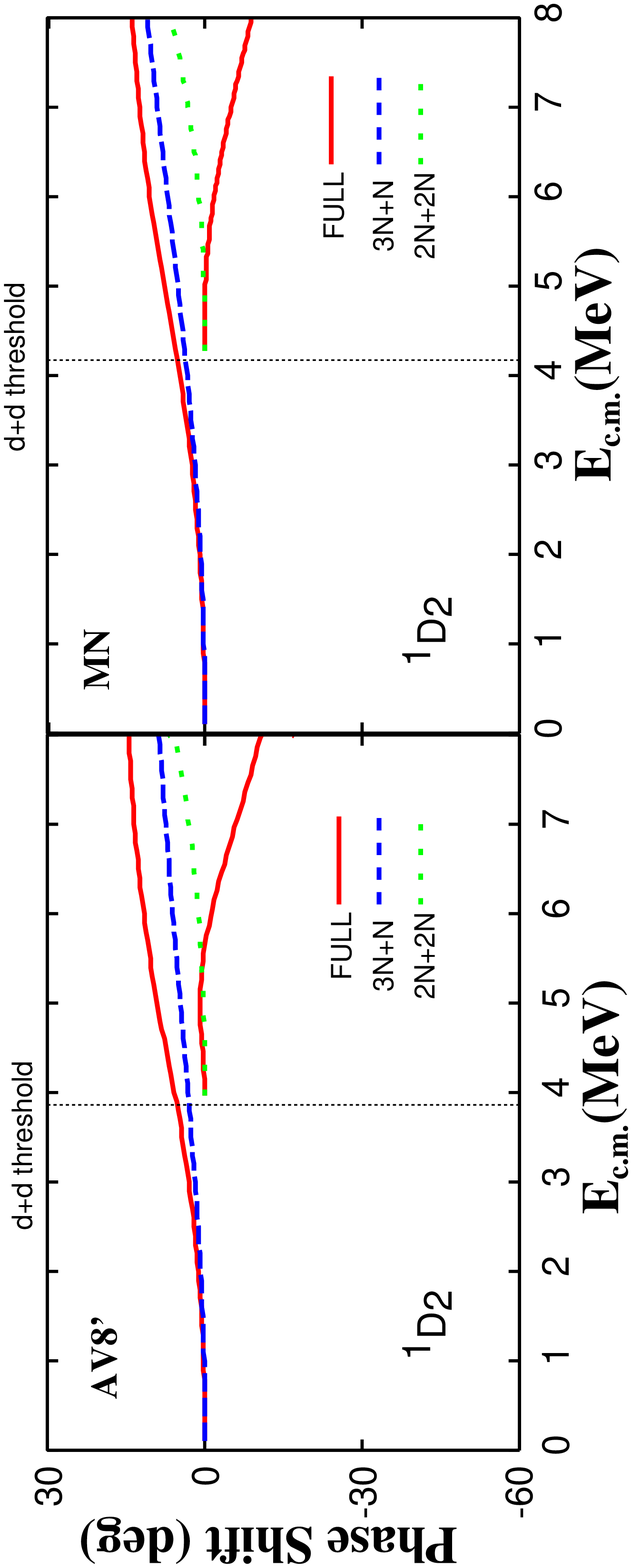}}
\caption{$^1D_2$ elastic-scattering phase shifts
with the AV8$^{\prime}$ (left) and MN (right) potentials.
The model space for the solid line is FULL. The $t$+$p$ phase shift
starts from the $t$+$p$ zero energy and the $d$+$d$ phase shift starts from the $d$+$d$ threshold.
The dashed line is the $t$+$p$ phase shift with only the 3$N$+$N$ channel
and the dotted line is 
the $d$+$d$ phase shift with only the 2$N$+2$N$ channel.
}
\label{fig:2+}
\end{figure}

As listed in Table~\ref{chan0}, there are four channels,
$^5S_2$, $^1D_2$, $^3D_2$ and $^5D_2$, for $J^{\pi}=2^+$ at 
energies around the $d$+$d$ threshold.
Among these states, we expect that the effect of the coupling between
the 3$N$+$N$ and 2$N$+2$N$ channels occurs most strongly
in $^1D_2$ as it appears
in all  physical channels. However, no sharp resonance is observed
in $^4$He up to 28\,MeV of excitation energy, so that
the coupling effect, if any, might be weaker than that observed in the
$^1S_0$ case.

Fig.~\ref{fig:2+} displays the $^1D_2$ elastic-scattering phase shifts
obtained in three types of calculations, $3N$+$N$ (dashed line),
$2N$+$2N$ (dotted line), and FULL (solid line).
The $t$+$p$ and $d$+$d$ phase shifts start from the $t$+$p$ ($E_{\rm c.m.}=0$) and
$d$+$d$ thresholds, respectively.
The phase shifts of the $3N$+$N$ and
$2N$+$2N$ calculations are both slightly positive, indicating a weak
attraction in the $t$+$p$ and $d$+$d$ interactions. In the FULL
calculation, the $t$+$p$ phase shift becomes more attractive
and the $d$+$d$ phase shift turns to be negative (repulsive).
The present FULL calculation reproduces the calculation of Ref.~\cite{hofmann08}
as expected.
Though the effect of the coupling is slightly larger in the AV8$^{\prime}$ potential
than in the MN potential, it is much less compared to
the case of the $^1S_0$ phase shift. This is understood as follows.
In the $^1D_2$ state, the main component of the
wave function is given by the $L=2$, $S=0$ state: Its probability
is the same as that of $^1S_0$, that is, 92\% in $t$+$p$ and
89\% in $d$+$d$. However, the
probability of finding the state with $L=0$, $S=2$, which
causes a strong tensor coupling, is more than
one order of magnitude smaller than in 
the case of $^1S_0$, namely 0.23\% in $t$+$p$ and 0.44\%
in $d$+$d$, respectively. The reason for this small percentage is 
that, to obtain $L=0$, the incoming $D$-wave in the
$^1D_2$ channel must couple with the $D$-components in
the clusters, but this coupling leads to several fragmented components
with different $L$ values.
This relatively weaker coupling of the tensor force explains the phase
shift behavior in Fig.~\ref{fig:2+}. 

\begin{figure}[ht]
\centerline{\includegraphics[width=6.0 cm,height=13.0 cm,angle=-90]{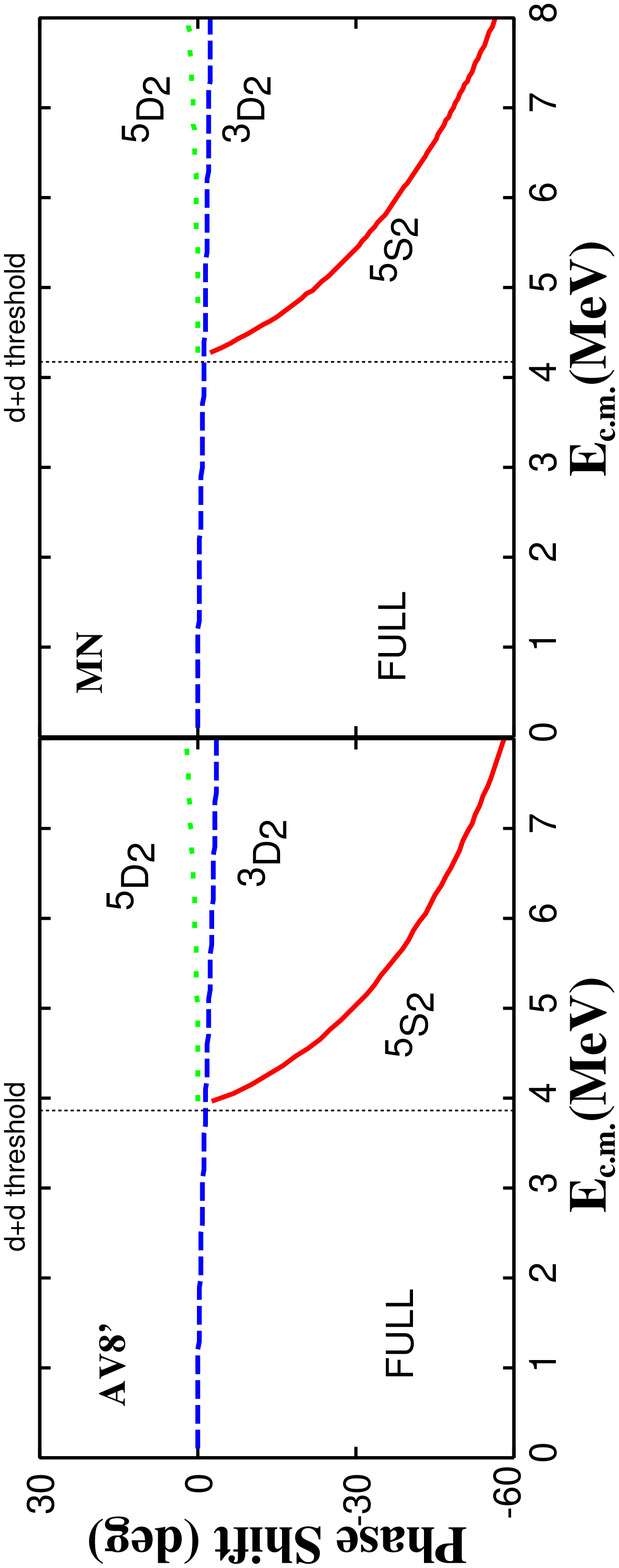}}
\caption{
Elastic scattering phase shifts of the 2$^+$ state obtained
in a FULL calculation
with the AV8$^{\prime}$ (left) and MN (right) potentials.
Solid line: $^{5}S_2$ $d$+$d$ phase shift, dotted line:
$^{5}D_2$ $d$+$d$ phase shift, dashed line: $^{3}D_2$ $t$+$p$
phase shift.}
\label{fig:2+full}
\end{figure}

In Fig.~\ref{fig:2+full} we plot the $t$+$p$ and
$d$+$d$ elastic-scattering phase shifts for other channels,
$^5S_2$ (solid line), $^3D_2$ (dashed line), and $^5D_2$ (dotted line).
We show only the  FULL result, because the phase shifts with the
truncated basis do not change visibly at the scale of the figure.
The obtained phase shifts are not that different between the AV8$^{\prime}$
and MN potentials, and also consistent with the previous calculation \cite{hofmann08}.
Thus, the effect of the distortion of the clusters is very small
for 2$^+$ except for $^1D_2$.

\begin{figure}[t]
\centerline{\includegraphics[width=6.0 cm,height=13.0 cm,angle=-90]{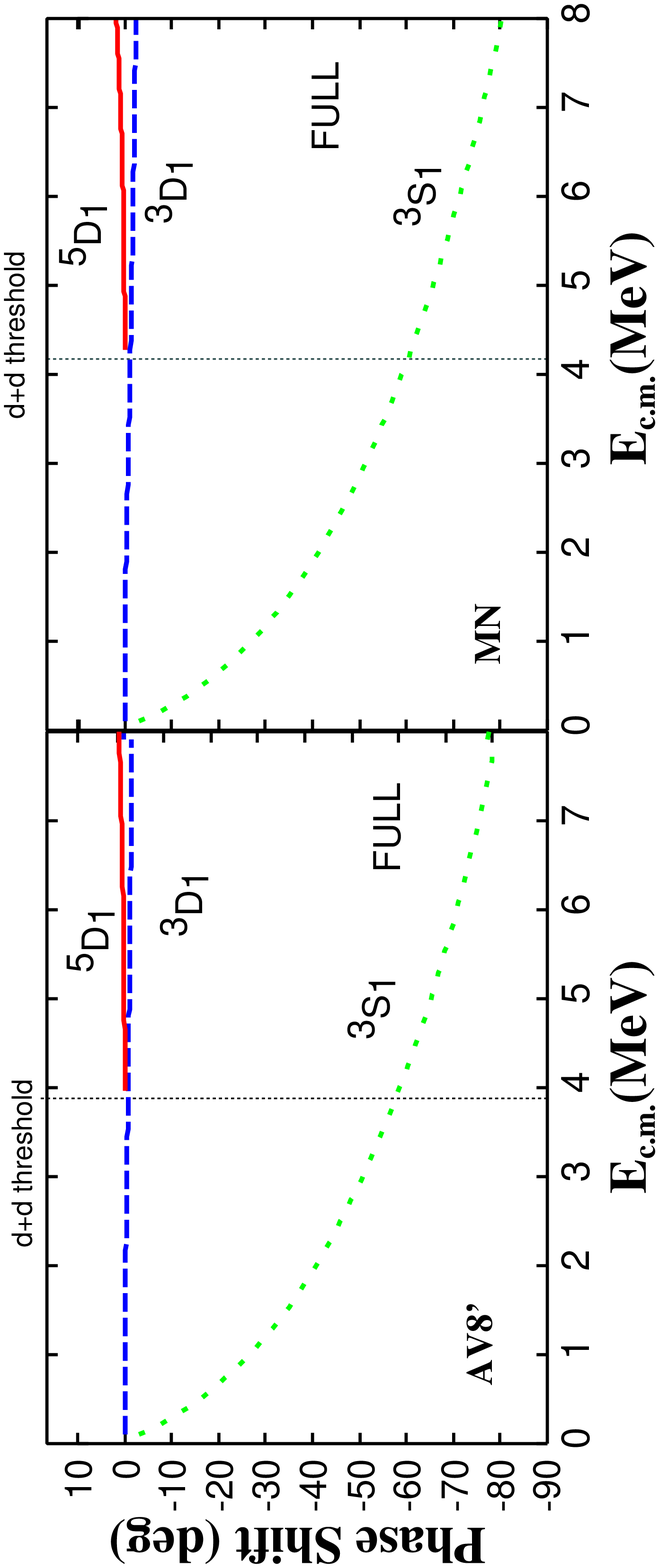}}
\caption{
Elastic scattering phase shifts of the 1$^+$ state in a
FULL calculation with the AV8$^{\prime}$ (left) and MN (right) potentials.
Solid line: $^{5}D_1$ $d$+$d$ phase shift, dotted line:
$^{3}S_1$ $t$+$p$ phase shift, dashed line: $^{3}D_1$ $t$+$p$
phase shift.}
\label{fig:1+}
\end{figure}

We have three channels for $J^{\pi}=1^+$,
$^5D_1$, $^3D_1$ and $^3S_1$. No sharp $1^+$ resonance of $^4$He is observed
experimentally up to 28\, MeV of excitation energy. Another theoretical
calculation neither predicts it \cite{horiuchi08}, so that
the coupling between the 2$N$+2$N$ and  3$N$+$N$ channels
is expected to be weak.
Fig.~\ref{fig:1+} exhibits the $t$+$p$ and $d$+$d$ elastic-scattering
phase shifts in the FULL calculation:
$^5D_1$ $d$+$d$ (solid line), $^3D_1$ $t$+$p$ (dashed line), and $^3S_1$
$t$+$p$ (dotted line).
Only the FULL result is displayed because the phase shift change in
other calculations is small. Both AV8$^{\prime}$ and MN potentials produce
phase shifts quite similar to each other. 

\begin{figure}[ht]
\centerline{\includegraphics[width=6.0 cm,height=13.0 cm,angle=-90]{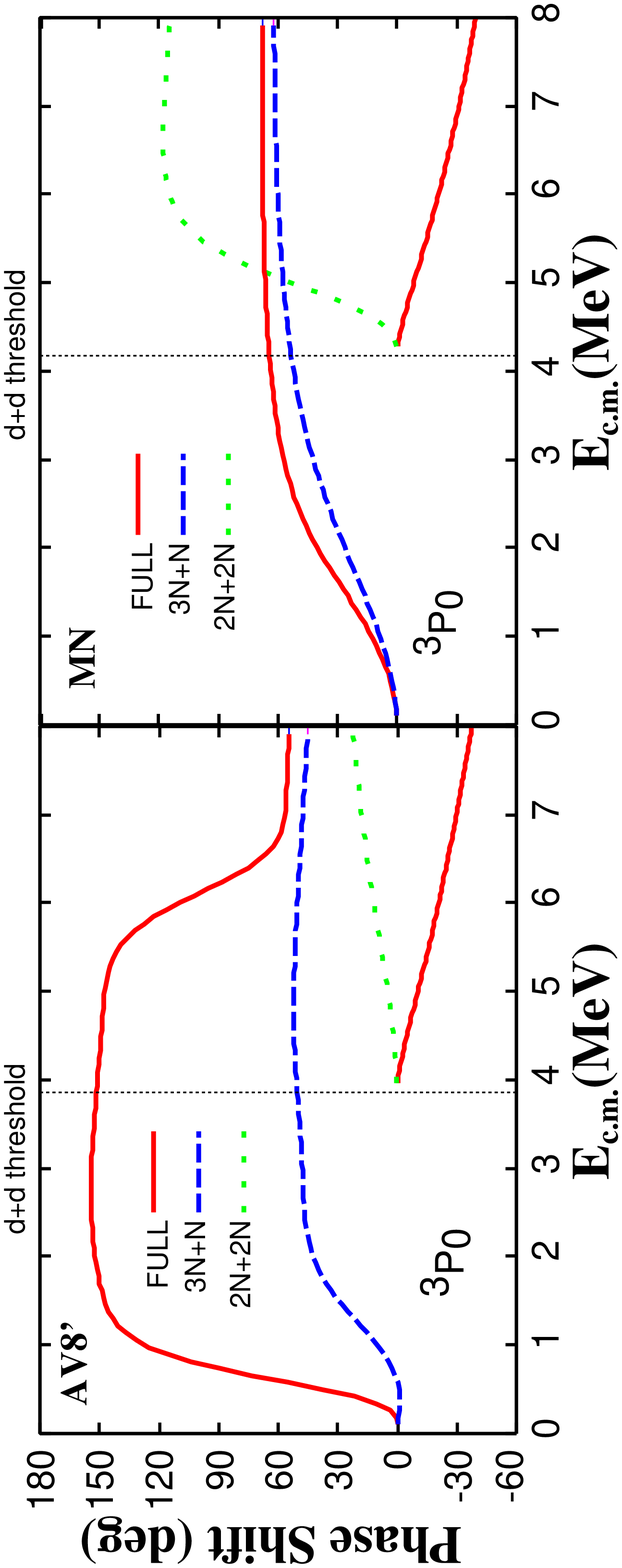}}
\caption{
$^3P_0$ elastic-scattering phase shifts calculated 
with the AV8$^{\prime}$ (left) and MN (right) potentials. See the caption of 
Fig.~\ref{fig:2+}. 
}
\label{fig:0-}
\end{figure}

\subsection{Negative parity phase shifts}

As seen from Table~\ref{chan0}, the main components of these negative 
parity states are considered to be $^3P_J$. 

We compare in Fig.~\ref{fig:0-} the $^3P_0$ elastic-scattering 
phase shifts calculated with the AV8$^{\prime}$ (left) and MN (right) potentials.
The truncated $3N$+$N$ (dashed line) and $2N$+$2N$ (dotted line) calculations 
are shown together with the FULL result (solid line). 
The $t$+$p$ phase shift of the $3N$+$N$ calculation 
is similar with both AV8$^{\prime}$ and 
MN potentials, while the $d$+$d$ phase shift of the $2N$+$2N$ calculation 
behaves quite differently between the two potentials: the $d$+$d$ phase shift is weakly 
attractive with AV8$^{\prime}$ but is very strongly attractive with MN. No typical 
resonance behavior shows up below the $d$+$d$ threshold, which is 
in contradiction to experiment. 
In the FULL model that combines both 3$N$+$N$ and 2$N$+2$N$ configurations,
however, the two potentials predict quite different phase shifts especially in 
the $t$+$p$ channel. The $t$+$p$ phase shift with AV8$^{\prime}$ becomes so 
attractive that it crosses $\pi/2$, indicating a resonance at 
about 1 MeV above the $t$+$p$ threshold. The $d$+$d$ phase shift changes 
sign from attractive to repulsive. The result based on the AV8$^{\prime}$
potential is thus consistent with experiment.
Furthermore, we reproduce the flat structure of the $^3P_0$ phase shift
around several MeV
above the $t$+$p$ threshold which was discussed as the coupling
to the $h$+$n$ channel \cite{hofmann08}.
On the other hand, the MN potential changes the $t$+$p$ phase shift only 
mildly and produces no sharp resonance behavior. The $d$+$d$ phase shift 
changes drastically to the repulsive side.

As seen in the above figure, the sharp 0$^-$ state appears provided 
a full model space with a realistic potential is employed. The mechanism 
to produce this resonance is unambiguously attributed to the tensor force
as discussed in Ref.~\cite{DGVR} for the realistic interaction G3RS \cite{tamagaki68}. According to it, 
the $0^-$ state consists of only two components, $L=S=1$ (95.5\%) and 
$L=S=2$ (4.5\%), ignoring a tiny component with 
$L=S=0$. The $L=S=2$ 
component arises from the coupling of the incoming $P$-wave with the 
$D$-states contained in the $3N$ and $d$ clusters. All the pieces
of the Hamiltonian but the tensor force have no coupling matrix element
between the two components. The uncoupled Hamiltonian thus 
gives a too high energy to accommodate a resonance. The tensor force, however, 
couples the two components very strongly, bringing down its energy to 
a right position.

\begin{figure}[ht]
\centerline{\includegraphics[width=6.0 cm,height=13.0 cm,angle=-90]{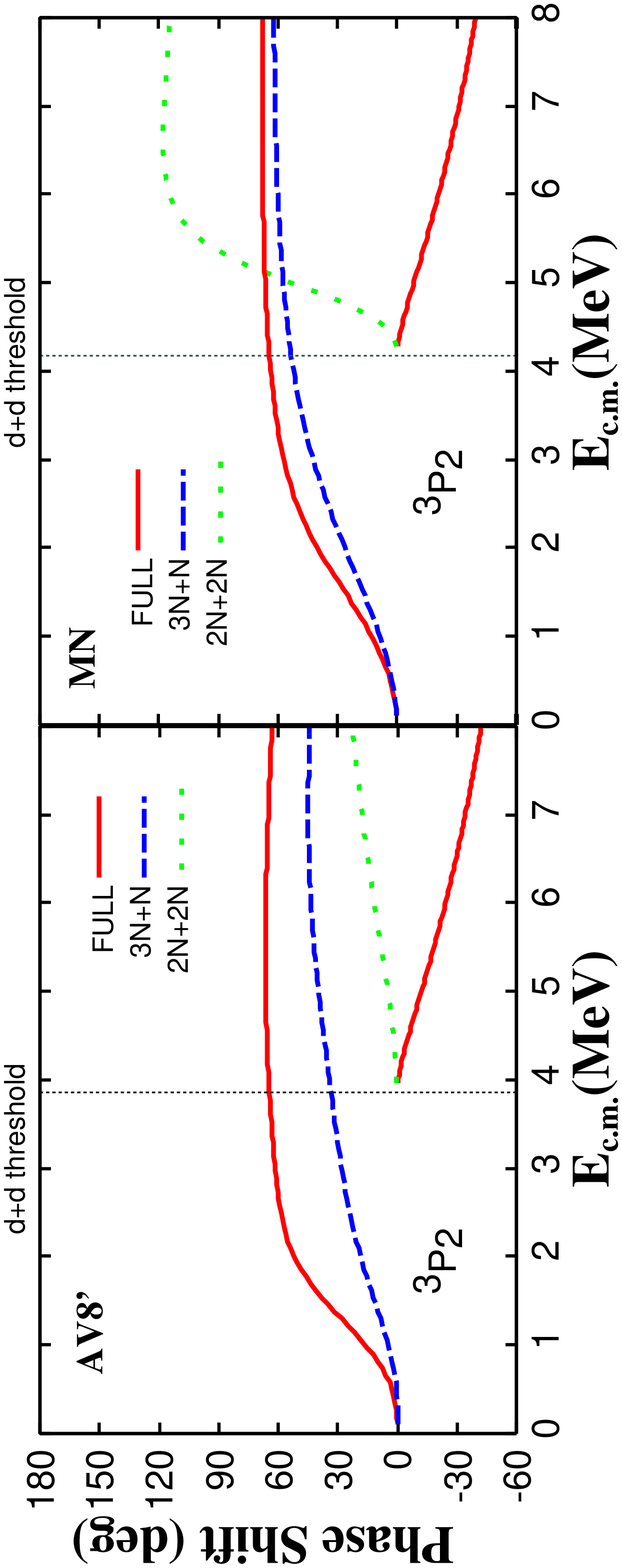}}
\caption{
$^3P_2$ elastic-scattering phase shifts calculated 
with the AV8$^{\prime}$ (left) and MN (right) potentials. See the caption of
 Fig.~\ref{fig:2+}.}
\label{fig:2-}
\end{figure}

The second lowest negative parity state has spin-parity 2$^-$. 
The physical channel for this state is only $^3P_2$ as seen in
Table~\ref{chan0}. Fig.~\ref{fig:2-} compares the 
$^3P_2$ elastic-scattering phase shifts in a manner similar 
to Fig.~\ref{fig:0-}.  The phase shift obtained with the 
MN potential is almost the same as the $^3P_0$ phase shift, which is 
consistent with the previous result~\cite{horiuchi08} that 
the energies of the negative parity states calculated with the 
MN potential are found to be degenerate. In the case of 
the AV8$^{\prime}$ potential, the $^3P_2$ phase shifts grows significantly in 
the FULL calculation, indicating a resonant behavior. The coupling effect between the $3N$+$N$ and 
$2N$+$2N$ channels is however much less compared to 
the $0^-$ state. This is because the incoming $P$-wave coupled to the 
$D$-states in the clusters gives rise to several $L$ values 
to produce the $2^-$ state and therefore the tensor coupling does not
concentrate sufficiently to produce a sharp resonance.

\begin{figure}[ht]
\centerline{\includegraphics[width=6.0 cm,height=13.0 cm,angle=-90]{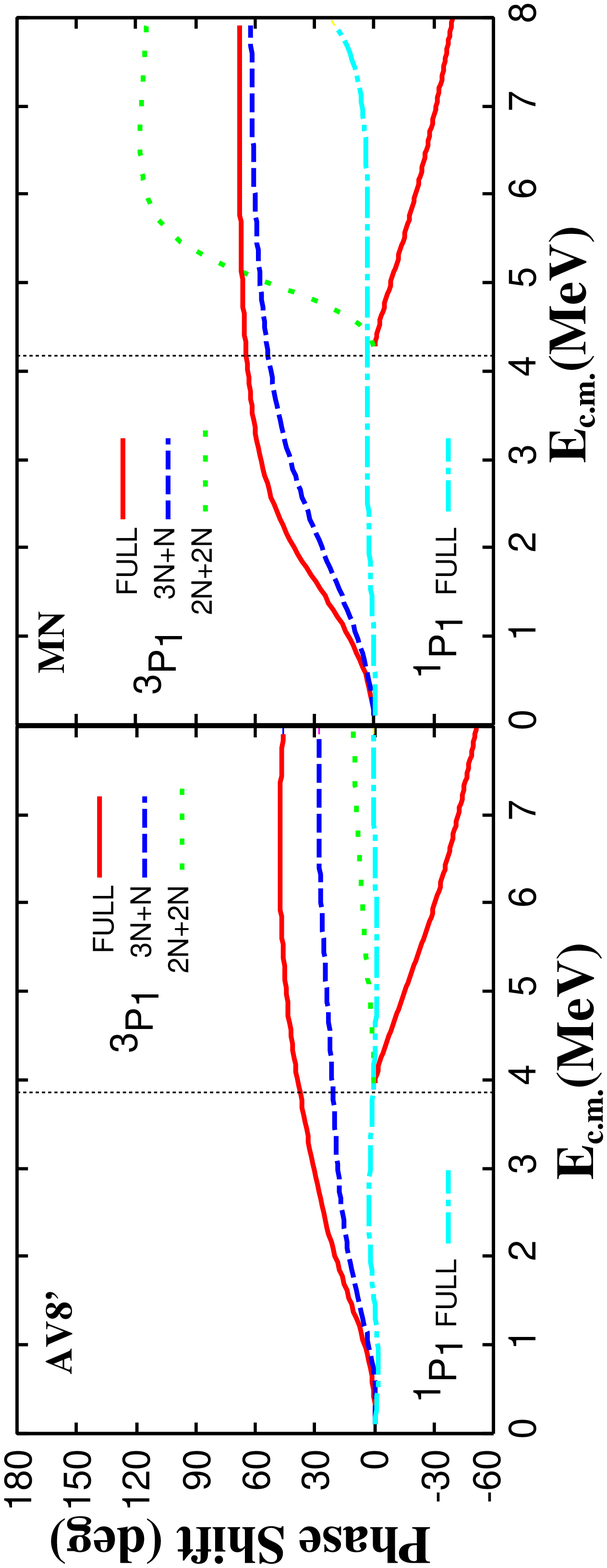}}
\caption{
$^3P_1$ and $^1P_1$ elastic-scattering phase shifts calculated 
with the AV8$^{\prime}$ (left) and MN (right) potentials. See the caption of
 Fig.~\ref{fig:2+}.} 
\label{fig:1-}
\end{figure}

Fig.~\ref{fig:1-} displays 
the $^3P_1$ and $^1P_1$ elastic-scattering phase shifts calculated 
with the AV8$^{\prime}$ (left) and MN (right) potentials. Note that 
no physical $d$+$d$ channel exists in the case of the $^1P_1$
state. Because both FULL and $3N$+$N$ calculations give almost the same 
phase shifts, only the FULL result is shown in the figure.  
The $^3P_1$ phase shift calculated with the MN potential is again 
almost the same as those of the $^3P_0$ and $^3P_2$ cases, supporting that 
the three negative parity states become almost degenerate.  
The $^3P_1$ elastic-scattering phase shift calculated with the 
AV8$^{\prime}$ potential is qualitatively similar to that of $^3P_2$. The
attractive nature of the $t$+$p$ phase shift becomes further weaker, and
to identify a resonance appears to be very hard. Even though it is
possible in some way, its width would be a few MeV, which is not in 
contradiction to experiment. The $^1P_1$ phase shifts are very
small in both AV8$^{\prime}$ and MN cases.

For the negative parity states, 
the FULL model with the AV8$^{\prime}$ potential gives results 
that are consistent with both experiment and the theoretical
calculation of Ref.~\cite{horiuchi08}. We have pointed out 
that the phase shift behavior 
reveals the importance of the tensor force particularly in the case 
of $0^-$. Its effect is often masked however by the coupling between the
$D$ states in the clusters and the incoming partial wave.

\begin{figure}[t]
\begin{center}
\epsfig{file=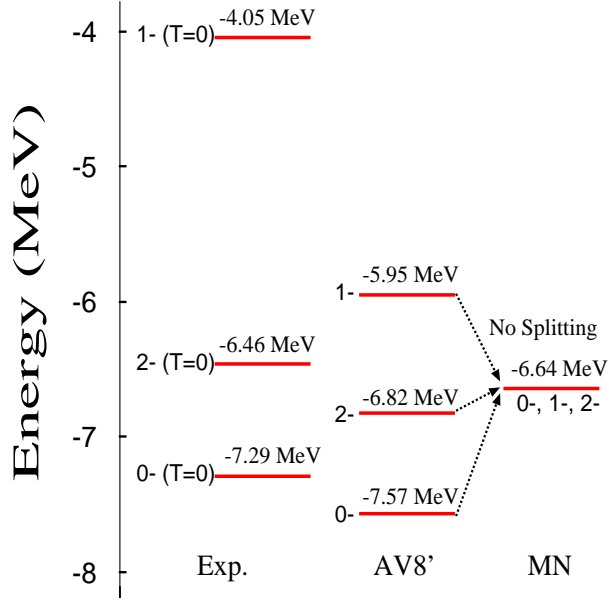,width=8.0cm,height=8.0cm}
\caption{Comparison of energies with respect to the four-nucleon
threshold for negative parity states (0$^-$, 1$^-$, 2$^-$)
between the experiment (left), AV8$^{\prime}$ (middle) and MN (right).}
\label{fig.13}
\end{center}
\end{figure}

In this subsection, we investigate the phase shifts of
the negative parity states which have dominant $T=0$ components.
In  Fig.~\ref{fig.13}, 
we represent three experimental negative parity $T=0$ energies (left).
The states are observed at $-7.29$ ($0^-$), $-6.46$ ($2^-$)
and $-4.05$ ($1^-$)\,MeV below the four-nucleon threshold \cite{tilley92}.
The former two are located below the $d$+$d$ threshold and their
widths are 0.84 and 2.01\,MeV, respectively, whereas the last one
is above the $d$+$d$ threshold and its width is fairly broad (6.1 MeV).
Here, we calculate these energies
as $-7.57$ ($0^-$), $-6.82$ ($2^-$)
and $-5.95$ ($1^-$)\,MeV, which are approximated by the half-value position
from the maximum phase shifts. 
The present calculation is not projected out to $T=0$, but the dominant
configurations of $t$+$p$, $h$+$n$ and $d$+$d$ elastic scattering are $T=0$.
Our calculated energies with AV8$^{\prime}$ reproduce the ordering of 
$0^-$, $2^-$ and $1^-$ (middle in Fig.~\ref{fig.13}).
The splitting between the two lower states $0^-$ and $2-$
is reproduced, but the experimental $1^-$ energy is higher than
the calculation. However, the determination of the energy
for such a high energy state with large decay width (6.1 MeV) 
is very difficult from both experimental and theoretical side,
and it usually has a large ambiguity.

This type of analysis was done by 
Horiuchi and Suzuki, who
applied the correlated Gaussian basis with two
global vectors to study the energy spectrum of $^4$He~\cite{horiuchi08}.
Because the results of Ref.~\cite{horiuchi08} are
based on approximate solutions that impose no proper resonance
boundary condition, it is interesting to see how the tensor force
changes the phase shifts in the negative parity states
as shown in this subsection.
These authors also found that the negative parity states with $T=0$
turn out to be almost degenerate when the MN potential that contains no tensor
force is employed. In the present calculation, three states ($0^-$, $1^-$, $2^-$)
completely degenerate at the same energy, $E=-6.64$ MeV (right), and the same
phase shift pattern (solid lines in Figs.~\ref{fig:0-},~\ref{fig:2-},~\ref{fig:1-}).
Thus we can expect to see a clear evidence for
the tensor force in the scattering involving the negative
parity states.

\section{Summary and conclusion}	
We have investigated the distortion of clusters appearing
in the low-energy $d$+$d$ and $t$+$p$ elastic scattering 
using a microscopic cluster model
with the triple global vector method.
We showed that the tensor interaction changes
the phase shifts very much by comparing a realistic interaction
and an effective interaction.
In the present $ab$-$initio$ type cluster model, 
the description of the cluster wave functions is extended 
from a simple (0$s$) harmonic-oscillator shell model to a few-body model.
To compare distortion effects of the clusters
with realistic and effective interactions,
we employed the AV8$^{\prime}$ potential as a realistic interaction and
the MN potential as an effective interaction.

For the realistic interaction,
the calculated $^1S_0$ phase shift shows that the
$t$+$p$ and $h$+$n$ channels strongly couple with the $d$+$d$ channel.
These channels are coupled because of the tensor interaction.
On the contrary, the coupling of these $3N$+$N$ channels plays a relatively
minor role for the case of the effective interaction because of
the absence of tensor term.
In other words, the $3N$+$N$ channels strongly affect the $d$+$d$
elastic phase shift with the realistic interaction, but not with the effective interaction.

For the 2$^+$ phase shifts, there is a $^1D_2$ component in all physical channels ($d$+$d$, 
$t$+$p$ and $h$+$n$). The coupling of the $2N$+$2N$ and $3N$+$N$ channels in $^5D_2$
is weaker than in $^1S_2$ because of a weaker tensor coupling
as discussed in section \ref{results}, and the calculated phase shifts
are very similar for the realistic and effective potentials.
For other positive parity cases,  the phase shift behavior
of the realistic and effective potentials are very similar,
and the coupling between the $2N$+$2N$ and $3N$+$N$ channels can be neglected
or is very small.
Furthermore, the tensor interaction makes the energy splitting
of the $0^-$, $2^-$ and $1^-$ negative parity states of $^4$He 
consistent with experiment.
No such splitting is however reproduced with the effective interaction.

We believe that the physical picture obtained in the large model space with
the realistic interaction should be close to the real physical situation.
It is needless to say that $ab$-$initio$ reaction calculations are very important
to understand the underlying reaction dynamics involving 
continuum states. 
Simpler calculations using effective interactions in the same framework,
as carried out in the present paper, are also meaningful 
because we can understand more clearly the effect of the tensor force by comparing both calculations.
The reaction calculations with the microscopic cluster model,
whose model space and interactions are restricted,
have been successfully applied to many heavier nuclei.
Therefore, it is instructive to see the difference from the realistic
interaction by employing a simple conventional
effective interaction as MN in the few-nucleon systems.

It will be quite interesting to see the importance 
of the tensor force in reaction observables of four nucleons. 
As a direct application of the present study the 
radiative capture reaction $d(d,\gamma)^4$He 
at energies of astrophysical interest is of prime importance. 
It is expected to take place predominantly via $E2$ 
transitions~\cite{santos85,langanke87,wachter88,arriaga91,carlson98}.
As is seen from Table \ref{chan0}, the two deuterons can approach 
each other in the $S$-wave only when $J^{\pi}$ is 
either 0$^+$ ($^1S_0$) or 2$^+$ ($^5S_2$). The former case is 
excluded  because a radiative capture reaction 
of $0^+ \to 0^+$ is forbidden in the lowest-order electromagnetic 
interaction, and hence the $E2$ transition should be predominant. 
If there were no tensor force present, the radiative capture 
would be suppressed near $E=0$ because neither $d$ nor $^4$He 
would have a $D$-wave component 
in contradiction with the flat behavior of 
the astrophysical $S$-factor~\cite{angulo99}. 
The tensor force strongly changes this story because 
it can couple $S$- and $D$-waves, bringing a 
significant amount of $D$-state probability in both $^4$He and
$d$. Details of this analysis will be reported elsewhere.
\\
\\
\noindent
{\bf Acknowledgment}\\
We thank Dr. R. Kamouni for helpful discussions based on his PhD thesis (in French).
This work presents research results of Bilateral Joint Research
Projects of the JSPS (Japan) and the FNRS (Belgium). Y.~S. is 
supported by a Grant-in-Aid for Scientific Research (No. 21540261).
This text presents research results of the Belgian program P6/23 on
interuniversity attraction poles initiated by the Belgian-state
Federal Services for Scientific, Technical and Cultural Affairs (FSTC). 
D.~B. and P.~D. also acknowledge travel support of the 
Fonds de la Recherche Scientifique Collective (FRSC).
The part of computational calculations were carried out in T2K-Tsukuba.

\bigskip

\appendix
\section{Definitions of recoupling coefficients}
\label{app.A}

We define an auxiliary coefficient $Z$ that appears in the coupling 
\begin{equation}
[[Y_a({\vi e}_1)[Y_b({\vi e}_1)Y_b({\vi e}_2)]_0]_a 
\to Z(ab)[Y_{a+b}({\vi e}_1)Y_b({\vi e}_2)]_a.
\end{equation}
By introducing a coefficient 
\begin{equation}
C(a b, c)=\sqrt{\frac{(2a+1)(2b+1)}
{4\pi(2c+1)}}\langle a\ 0\ b\ 0 \vert c \ 0\rangle \\
\end{equation}
for the coupling 
$[Y_a({\vi e}_1)Y_b({\vi e}_1)]_{c}=C(ab, c)Y_{c}({\vi e}_1)$, we 
can express $Z$ as 
\begin{equation}
Z(ab)=\sqrt{\frac{2(a+b)+1}{(2a+1)(2b+1)}}C(ab,a+b)=
\frac{1}{\sqrt{4\pi}}\langle a\ 0\ b\ 0 \vert a+b \ 0\rangle.
\label{def.z}
\end{equation}
Note that $C(ab,c)$ vanishes unless $a+b+c$ is even. 

The coefficients that appear in Sect.~\ref{sect.4} are given as 
follows: 
\begin{eqnarray}
&&{\hspace{-5mm}}X(a\ b\ c)=Z(ab)Z(ac)C(bc, b+c)U(a\!+\!c\ c\ a\!+\!b\ b;\ a\ b\!+\!c),\\
&&{\hspace{-5mm}}R_3(a\ b\ c)=Z(ab)Z(a\!+\!b \ c),\\
&&{\hspace{-5mm}}W(a\ b\ c\ q\ Q,\ a'\ b'\ c'\ q'\ Q',\ \ell \ \ell')\nonumber\\
&&{\hspace{-5mm}}=
\left[\begin{array}{ccc}
q &c & Q \\
q'&c' & Q'\\
\ell'&c\!+\!c' &\ell\\
\end{array}\right]
\left[\begin{array}{ccc}
a &b & q\\
a' &b' & q'\\
a\!+\!a'&b\!+\!b' &\ell'\\
\end{array}\right]C(aa', a\!+\!a')C(bb', b\!+\!b')C(cc', c\!+\!c').
\end{eqnarray}

\section{Matrix elements for various operators}
The purpose of this appendix is to collect formulas 
for various matrix elements. The main procedure to derive the 
formulas is sketched in Sect.~\ref{sect.4}. More details for the case 
of two global vectors are given in Ref.~\cite{DGVR}.
 
\label{app.B}
\subsection{Kinetic energy}
Let ${\vi \pi}_j$ denote the momentum operator conjugate to 
${\vi x}_j$, ${\vi \pi}_j=-i\hbar \frac{\partial}{\partial {\vi x}_j}$. 
The total kinetic energy operator for the $N$-nucleon system with its 
center of mass kinetic energy being subtracted takes the form 
\begin{equation}
\sum_{i=1}^N \frac{{\vi p}_i^2}{2m}-\frac{{\vi \pi}_N^2}{2Nm}
=\frac{1}{2}\widetilde{\vi \pi}\Lambda{\vi \pi},
\end{equation}
where ${\vi \pi}_N=\sum_{i=1}^N{\vi p}_i$ is the total momentum, 
$\widetilde{\vi \pi}=({\vi \pi}_1,{\vi \pi}_2,\ldots,{\vi \pi}_{N-1})$, and 
$\Lambda$ is an $(N-1)\times(N-1)$ symmetric mass matrix. Defining $N-1$-dimensional column vectors $\Gamma_i$ as 
\begin{eqnarray}
& &\Gamma_i=A'B^{-1}u_i\ \ \ \ \ \ (i=1,2,3),\nonumber \\
& &\Gamma_i=-AB^{-1}u_i\ \ \ \ \ (i=4,5,6)
\label{def.gamma}
\end{eqnarray}
and an $(N-1)\times(N-1)$ matrix $Q$
\begin{equation}
Q_{ij}=2\widetilde{\Gamma_i}\Lambda \Gamma_j,
\end{equation}
we can calculate the matrix element for the kinetic energy through the 
overlap matrix element
\begin{eqnarray}
\langle F'\vert \frac{1}{2}\widetilde{\vi \pi}\Lambda{\vi \pi} \vert F \rangle
= \frac{\hbar^2}{2}\left(R-\sum_{i<j}Q_{ij}\frac{\partial}{\partial \rho_{ij}}\right)
\left<F'\vert F \right>,
\end{eqnarray}
where 
\begin{equation}
R=3{\rm Tr}(B^{-1}A'\Lambda A).
\end{equation}
The $\rho_{ij}$ values are defined in Eq.~(\ref{def.rho}). 

\subsection{$\delta$-function}
A two-body interaction $V({\vi r}_i-{\vi r}_j)$ can be expressed as 
\begin{equation}
V({\vi r}_i-{\vi r}_j)=\int d{\vi r} V({\vi r})\ \delta({\vi r}_i-{\vi r}_j-{\vi r}).
\end{equation}
Once the matrix element of $\delta({\vi r}_i-{\vi r}_j-{\vi r})$ 
is obtained, the matrix element of the interaction is calculated by integrating over ${\vi r}$ the $\delta$-function matrix element weighted 
with the form factor $V({\vi r})$. Similarly, for a one-body operator 
\begin{equation}
D({\vi r}_i-{\vi x}_N)=\int d{\vi r} D({\vi r})\ \delta({\vi r}_i-{\vi x}_N-{\vi r}),
\end{equation}
its matrix element can be obtained from 
that of the $\delta$-function. Because both 
${\vi r}_i-{\vi r}_j$ and ${\vi r}_i-{\vi x}_N$ can be expressed in terms of 
a linear combination of the relative coordinate ${\vi x}_i$, it is 
enough to calculate the matrix element of 
$\delta(\widetilde{w}{\vi x}-{\vi r})$, 
where $\widetilde{w}=(w_1,w_2,\ldots,
w_{N-1})$ is a combination constant to express ${\vi r}_i-{\vi r}_j$ 
or ${\vi r}_i-{\vi x}_N$. 

The matrix element of the $\delta$-function is given by 
\begin{eqnarray}
& &{\hspace{-1cm}}\left<F'\vert \delta(\widetilde{w}{\vi x}-{\vi r}) \vert F \right>
\nonumber \\
&=&\left(\frac{(2\pi)^{N-1}}{\mbox{det} B}\right)^{3/2}
\left(\prod_{i=1}^6 B_{L_i}\right)
\frac{(-1)^{L_1+L_2+L_3}\sqrt{2L+1}}{\sqrt{2L'+1}}
\ \left(\frac{c}{2\pi}\right)^{3/2}{\rm e}^{-\frac{1}{2}cr^2} \nonumber\\
&\times & \sum_{\kappa \mu} \langle LM\kappa \mu| L'M'\rangle 
Y_{\kappa \mu}^*(\widehat{\vi r}) 
 \sum_{p_i} \left(\prod_{i=1}^6 (-c\gamma_i r)^{p_{i}}
\frac{ \sqrt{2p_{i}+1}}{B_{p_{i}}} \right)
\nonumber \\
&\times& \sum_{\ell_{12}\ell_{45}\ell \ell' \overline{L}_{12} 
\overline{L}_{45} \overline{L}} 
\frac{(-1)^{\ell+\ell'}}{\sqrt{(2\ell+1)(2\overline{L}+1)}}U(L\overline{L}\kappa
\ell';\ell L')\ 
\overline{O}(p_i ; \ell_{12} \ell_{45} \ell \ell' \kappa )
\nonumber \\
&\times& 
W(p_1 p_2 p_3 \ell_{12} \ell, L_1\!-\!p_1\ L_2\!-\!p_2\ L_3\!-\!p_3\ 
\overline{L}_{12}\overline{L}, L L_{12})\nonumber \\
&\times& 
W(p_4 p_5 p_6 \ell_{45} \ell', L_4\!-\!p_4\ L_5\!-\!p_5\ L_6\!-\!p_6\ 
\overline{L}_{45}\overline{L}, L' L_{45})\nonumber \\
&\times& 
\sum_{n_{ij}}\left(\prod_{i<j}^6 (-\overline{\rho}_{ij})^{n_{ij}}
\frac{ \sqrt{2n_{ij}+1}}{B_{n_{ij}}}\right)
\nonumber \\
&\times& 
O(n_{ij}; L_1\!-\!p_1\ L_2\!-\!p_2\ L_3\!-\!p_3\ L_4\!-\!p_4, 
L_5\!-\!p_5\ L_6\!-\!p_6,\overline{L}_{12}\overline{L}_{45}\overline{L}),
\label{me.del}
\end{eqnarray}
with
\begin{eqnarray}
& &c=(\widetilde{w}B^{-1}w)^{-1},\ \ \ \ \ \gamma_i=\widetilde{w}B^{-1}u_i,
\ \ \ \ \ \overline{\rho}_{ij}=\rho_{ij}-c\gamma_i\gamma_j.
\end{eqnarray}
The summation over non-negative integers $n_{ij}$ and $p_i$ is restricted by 
the following equations
\begin{eqnarray}
&&n_{12}+n_{13}+n_{14}+n_{15}+n_{16}+p_1=L_1, \nonumber\\
&&n_{12}+n_{23}+n_{24}+n_{25}+n_{26}+p_2=L_2, \nonumber\\
&&n_{13}+n_{23}+n_{34}+n_{35}+n_{36}+p_3=L_3, \nonumber\\
&&n_{14}+n_{24}+n_{34}+n_{45}+n_{46}+p_4=L_4, \nonumber\\
&&n_{15}+n_{25}+n_{35}+n_{45}+n_{56}+p_5=L_5, \nonumber\\
&&n_{16}+n_{26}+n_{36}+n_{46}+n_{56}+p_6=L_6. 
\end{eqnarray}
Here $\overline{O}(p_i ; \ell_{12} \ell_{45} \ell \ell' \kappa )$ 
is defined as a coefficient that appears in the coupling of a product of six terms
\begin{eqnarray}
& &{\hspace{-1cm}}\prod_{i=1}^6 [Y_{p_i}({\vi e}_i)Y_{p_i}(\widehat{\vi r})]_{00}
= \sum_{\ell_{12}\ell_{45}\ell \ell' \kappa}\overline{O}(p_i ; \ell_{12} \ell_{45} \ell \ell' \kappa)
\nonumber \\
&\times& 
[[[[Y_{p_1}({\vi e}_1)Y_{p_2}({\vi e}_2)]_{\ell_{12}}Y_{p_3}({\vi e}_3)]_{\ell}\ 
[[Y_{p_4}({\vi e}_4)Y_{p_5}({\vi e}_5)]_{\ell_{45}}Y_{p_6}({\vi e}_6)]_{\ell'}]_{\kappa}\ 
Y_{\kappa}(\widehat{\vi r})]_{00},
\end{eqnarray}
and it is given by 
\begin{eqnarray}
& &\overline{O}(p_i ; \ell_{12} \ell_{45} \ell \ell' \kappa )
\nonumber \\
& &\quad=\sqrt{\frac{2\kappa+1}{\prod_{i=1}^6(2p_i+1)}}
C(p_1 p_2,\ell_{12})C(\ell_{12} p_3,\ell)C(p_4 p_5,\ell_{45})C(\ell_{45} p_6,\ell')
C(\ell \ell', \kappa).
\end{eqnarray}

The ${\vi r}$-dependence of the matrix element~(\ref{me.del}) is 
\begin{equation}
{\rm e}^{-\frac{1}{2}cr^2} r^{p_1+p_2+p_3+p_4+p_5+p_6} Y_{\kappa \mu}^*(\widehat{\vi r}). 
\end{equation}
For a central interaction, $V({\vi r})$ is a scalar function, and the sum over 
$\kappa$ in Eq.~(\ref{me.del}) is 
limited to 0. For a tensor interaction, the angular dependence of $V({\vi r})$ is 
proportional to $Y_2(\hat{\vi r})$, and $\kappa$ is limited to 2.  
The electric multipole operator is a special case of one-body operator, 
so that one can make use of the formula~(\ref{me.del}) to calculate its
matrix element. More explicitly, we give 
the matrix element of  $V(|\widetilde{w}{\vi x}|) 
Y_{\kappa \mu}(\widehat{\widetilde{w}{\vi x}})$ that includes all the 
cases mentioned above: 
\begin{eqnarray}
& &{\hspace{-1cm}}\left<F'\vert V(|\widetilde{w}{\vi x}|) 
Y_{\kappa \mu}(\widehat{\widetilde{w}{\vi x}}) \vert F \right>
\nonumber \\
&=&\left(\frac{(2\pi)^{N-1}}{\mbox{det} B}\right)^{3/2}
\left(\prod_{i=1}^6 B_{L_i}\right)
\frac{(-1)^{L_1+L_2+L_3}\sqrt{2L+1}}{\sqrt{2L'+1}} \nonumber\\
&\times & \langle LM\kappa \mu| L'M'\rangle 
 \sum_{p_i} \left(\prod_{i=1}^6 (-\gamma_i )^{p_{i}}
\frac{ \sqrt{2p_{i}+1}}{B_{p_{i}}} \right){\cal I}^{(2)}_{p_1+p_2+p_3+p_4+p_5+p_6}(c)
\nonumber \\
&\times& \sum_{\ell_{12}\ell_{45}\ell \ell' \overline{L}_{12} 
\overline{L}_{45} \overline{L}} 
\frac{(-1)^{\ell+\ell'}}{\sqrt{(2\ell+1)(2\overline{L}+1)}}U(L\overline{L}\kappa
\ell';\ell L')\ 
\overline{O}(p_i ; \ell_{12} \ell_{45} \ell \ell' \kappa )
\nonumber \\
&\times& 
W(p_1 p_2 p_3 \ell_{12} \ell, L_1\!-\!p_1\ L_2\!-\!p_2\ L_3\!-\!p_3\ 
\overline{L}_{12}\overline{L}, L L_{12})\nonumber \\
&\times& 
W(p_4 p_5 p_6 \ell_{45} \ell', L_4\!-\!p_4\ L_5\!-\!p_5\ L_6\!-\!p_6\ 
\overline{L}_{45}\overline{L}, L' L_{45})\nonumber \\
&\times& 
\sum_{n_{ij}}\left(\prod_{i<j}^6 (-\overline{\rho}_{ij})^{n_{ij}}
\frac{ \sqrt{2n_{ij}+1}}{B_{n_{ij}}}\right)
\nonumber \\
&\times& 
O(n_{ij}; L_1\!-\!p_1\ L_2\!-\!p_2\ L_3\!-\!p_3\ L_4\!-\!p_4, 
L_5\!-\!p_5\ L_6\!-\!p_6,\overline{L}_{12}\overline{L}_{45}\overline{L}),
\end{eqnarray}
with the integral of the potential form factor 
\begin{eqnarray}
{\cal I}^{(m)}_n(c)=\left(\frac{c}{2\pi}\right)^{3/2} c^n\int_0^{\infty} dr \,r^{n+m} V(r) {\rm e}^{-\frac{1}{2}cr^2}. 
\end{eqnarray}
In case $V(r)$ takes the form of $r^q {\rm e}^{-\rho r^2-\rho'r}$ $(q
\geq -m)$, 
the integral ${\cal I}^{(m)}_n(c)$ can be obtained analytically, giving 
a closed form for the matrix element. 

It should be noted that the matrix element for a special class of a three-body force can be 
evaluated with ease. For example, if the radial part of the three-body 
force has a form 
\begin{equation}
V_{TNF}=\exp(-\rho_1({\vi r}_i-{\vi r}_j)^2-
\rho_2({\vi r}_j-{\vi r}_k)^2 
-\rho_3({\vi r}_k-{\vi r}_i)^2),
\end{equation}
the exponent can be rewritten as $-\widetilde{\vi x}\Omega {\vi x}$ 
with an $(N-1)\times(N-1)$ symmetric matrix  
$\Omega=\rho_1 w_{ij}\widetilde{w_{ij}}+\rho_2
w_{jk}\widetilde{w_{jk}}+\rho_3 w_{ki}\widetilde{w_{ki}}$,
where 
$w_{ij}$, $w_{jk}$  and $w_{ki}$ are defined by
${\vi r}_i-{\vi r}_j=\widetilde{w_{ij}}{\vi x}$, 
${\vi r}_j-{\vi r}_k=\widetilde{w_{jk}}{\vi x}$ and
${\vi r}_k-{\vi r}_i=\widetilde{w_{ki}}{\vi x}$.
Thus the matrix element 
reduces to that of the overlap with $A$ being replaced with 
$A+2\Omega$
\begin{equation}
\langle F'\vert V_{TNF} \vert F \rangle
= \langle F'\vert  F_{L_1 L_2(L_{12}) L_3 LM}(u_1, u_2, u_3, 
A+2\Omega,{\vi x}) \rangle.
\end{equation}

\subsection{Spin-orbit potential}
The spatial form of a spin-orbit interaction reads
\begin{equation}
V(|{\vi r}_i-{\vi r}_j|)(({\vi r}_i-{\vi r}_j)\times \frac{1}{2}({\vi p}_i-{\vi p}_j))_{\mu},
\end{equation}
where $({\vi a}\times{\vi b})_{\mu}$ $(\mu=0, \pm 1)$ stands for the 
$\mu$th component of a vector product of ${\vi a}$ and ${\vi b}$. 
As in the $\delta$-function matrix element, the spin-orbit potential is written as 
\begin{equation}
V(|\widetilde{w}{\vi x}|) (\widetilde{w}{\vi x}\times \widetilde{\zeta}{\vi \pi})_{\mu},
\end{equation}
where $\frac{1}{2}({\vi p}_i-{\vi p}_j)$ is expressed in terms of the momentum operators ${\vi \pi}$, $\widetilde{\zeta}{\vi \pi}=\sum_{i=1}^{N-1} \zeta_i {\vi \pi}_i$.

The matrix element of the spin-orbit potential is given by 
\begin{eqnarray}
& &{\hspace{-1cm}}\left<F'\vert V(|\widetilde{w}{\vi x}|) (\widetilde{w}{\vi x}\times \widetilde{\zeta}{\vi \pi})_{\mu} \vert F \right>
\nonumber \\
&=&\frac{4\pi \sqrt{2} \hbar}{3}\left(\frac{(2\pi)^{N-1}}{\mbox{det} B}\right)^{3/2}
\left(\prod_{i=1}^6 B_{L_i}\right)
\frac{(-1)^{L_1+L_2+L_3}\sqrt{2L+1}}{\sqrt{2L'+1}}
\nonumber\\
&\times & \langle LM 1 \mu| L'M'\rangle 
 \sum_{p_i} \left(\prod_{i=1}^6 (-\gamma_i )^{p_{i}}
\frac{ \sqrt{2p_{i}+1}}{B_{p_{i}}} \right){\cal I}^{(3)}_{p_1+p_2+p_3+p_4+p_5+p_6}(c)
\nonumber \\
&\times& \sum_{\ell_{12}\ell_{45}\ell \ell' \overline{\ell}_{12}
 \overline{\ell}_{45} \overline{\ell}\ \overline{\ell}'
 \overline{L}_{12} \overline{L}_{45} \overline{L} } 
\frac{(-1)^{\overline{\ell}+\overline{\ell}'}}
{\sqrt{(2\overline{\ell}+1)(2\overline{L}+1)}}
U(L\overline{L}1 \overline{\ell}'; \overline{\ell} L')\ 
\overline{O}(p_i ; \ell_{12} \ell_{45} \ell \ell' 1)
\nonumber \\
&\times&
\sum_{k=1}^6 (\widetilde{\zeta}\Gamma)_k \ T_k(p_i,\ell_{12}\ell_{45}\ell \ell', \overline{\ell}_{12} \overline{\ell}_{45} \overline{\ell}\ \overline{\ell}')
\sum_{n_{ij}}\left(\prod_{i<j}^6 (-\overline{\rho}_{ij})^{n_{ij}}
\frac{ \sqrt{2n_{ij}+1}}{B_{n_{ij}}}\right)
\nonumber \\
&\times&
O(n_{ij}; L_1\!-\!p_1^k\ L_2\!-\!p_2^k\ L_3\!-\!p_3^k\ L_4\!-\!p_4^k\  
L_5\!-\!p_5^k\ L_6\!-\!p_6^k,\overline{L}_{12}\overline{L}_{45}\overline{L}),
\end{eqnarray}
where ${p}_i^k$ $(k=1,2,\ldots,6)$ is 
\begin{equation}
{p}_i^k=p_i+\delta_{ik},
\end{equation}
and where the non-negative integers $n_{ij}$ and $p_i$ 
are constrained to satisfy the equations 
\begin{eqnarray}
&&n_{12}+n_{13}+n_{14}+n_{15}+n_{16}+p_1^k=L_1, \nonumber\\
&&n_{12}+n_{23}+n_{24}+n_{25}+n_{26}+p_2^k=L_2, \nonumber\\
&&n_{13}+n_{23}+n_{34}+n_{35}+n_{36}+p_3^k=L_3, \nonumber\\
&&n_{14}+n_{24}+n_{34}+n_{45}+n_{46}+p_4^k=L_4, \nonumber\\
&&n_{15}+n_{25}+n_{35}+n_{45}+n_{56}+p_5^k=L_5, \nonumber\\
&&n_{16}+n_{26}+n_{36}+n_{46}+n_{56}+p_6^k=L_6. 
\end{eqnarray}
The symbol $(\widetilde{\zeta}\Gamma)_k$ stands for the factor 
\begin{equation}
(\widetilde{\zeta}\Gamma)_k = \sum_{i=1}^6 \zeta_i (\Gamma_k)_i,
\end{equation}
where $(\Gamma_k)_i$ is the $i$th element of the column vector 
$\Gamma_k$ defined in Eq.~(\ref{def.gamma}). The coefficient $T_k$ 
appears in the coupling 
\begin{eqnarray}
& &[Y_1({\vi e}_k)\ [[[Y_{p_1}({\vi e}_1)Y_{p_2}({\vi e}_2)]_{\ell_{12}}Y_{p_3}({\vi e}_3)]_{\ell}\ 
[[Y_{p_4}({\vi e}_4)Y_{p_5}({\vi e}_5)]_{\ell_{45}}Y_{p_6}({\vi e}_6)]_{\ell'}]_{1}]_{1\mu}
\nonumber \\
& &\to \sum_{\overline{\ell}_{12} \overline{\ell}_{45} \overline{\ell}\ \overline{\ell}'}T_k(p_i,\ell_{12}\ell_{45}\ell \ell', \overline{\ell}_{12} \overline{\ell}_{45} \overline{\ell}\ \overline{\ell}')
\nonumber \\
& &\qquad \times 
[[[Y_{{p}_1^k}({\vi e}_1)Y_{{p}_2^k}({\vi e}_2)]_{\overline{\ell}_{12}}Y_{{p}_3^k}({\vi e}_3)]_{\overline{\ell}}\ 
[[Y_{{p}_4^k}({\vi e}_4)Y_{{p}_5^k}({\vi e}_5)]_{\overline{\ell}_{45}}Y_{{p}_6^k}({\vi e}_6)]_{\overline{\ell}'}]_{1\mu}.
\end{eqnarray}
The coefficients $T_k(p_i,\ell_{12}\ell_{45}\ell \ell', \overline{\ell}_{12} \overline{\ell}_{45} \overline{\ell}\ \overline{\ell}')$ are given below:
\begin{eqnarray}
& &T_1=U(1\ell 1 \ell'; \overline{\ell}1)
U(1 \ell_{12} \overline{\ell}p_3; \overline{\ell}_{12} \ell) 
U(1 p_1 \overline{\ell}_{12} p_2; p_1\!+\!1\ \ell_{12}) 
C(1 p_1; p_1\!+\!1)
\nonumber \\
& &T_2=-(-1)^{\ell_{12}+\overline{\ell}_{12}}
U(1\ell 1 \ell'; \overline{\ell}1)
U(1\ell_{12}\overline{\ell}p_3; \overline{\ell}_{12} \ell)
U(1 p_2 \overline{\ell}_{12} p_1; p_2\!+\!1\ \ell_{12}) 
C(1 p_2; p_2\!+\!1)
\nonumber \\
& &T_3=-(-1)^{\ell+\overline{\ell}}
U(1\ell 1 \ell'; \overline{\ell}1)
U(1p_3 \overline{\ell} \ell_{12} ; p_3\!+\!1 \ell) 
C(1 p_3; p_3\!+\!1)
\nonumber \\
& &T_4=(-1)^{\ell'+\overline{\ell}'}
U(1\ell' 1 \ell; \overline{\ell}'1)
U(1 \ell_{45} \overline{\ell}' p_4; \overline{\ell}_{45} \ell') 
U(1 p_4 \overline{\ell}_{45} p_5; p_4\!+\!1 \ \ell_{45}) 
C(1 p_4; p_4\!+\!1)
\nonumber \\
& &T_5=-(-1)^{\ell'+\overline{\ell}'+\ell_{45}+\overline{\ell}_{45}}
U(1\ell' 1 \ell; \overline{\ell}'1)
U(1\ell_{45} \overline{\ell}' p_6; \overline{\ell}_{45}\ell')
U(1 p_5 \overline{\ell}_{45} p_4; p_5\!+\!1\  \ell_{45}) 
C(1 p_5; p_5\!+\!1)
\nonumber \\
& &T_6=-U(1\ell' 1 \ell; \overline{\ell}'1)
U(1 p_6 \overline{\ell}' \ell_{45} ; p_6\!+\!1\ \ell') 
C(1 p_6; p_6\!+\!1).
\end{eqnarray}

\section{Momentum representation of correlated Gaussian basis}
\label{app.C}

The Fourier transform of the correlated Gaussian function~(\ref{cgtgv}) 
defines the corresponding basis function in momentum space. 
The momentum space function is useful to evaluate those 
matrix elements which depend on the momentum operator~\cite{DGVR}. 
Suppose that 
we want to evaluate the matrix element of a two-body 
operator $V({\vi p}_i-{\vi p}_j)$ 
or a one-body operator $D({\vi p}_i-\frac{1}{N}{\vi \pi}_N)$. 
Obviously evaluating the matrix element can be 
done more easily in momentum space. For this purpose we need to 
obtain the Fourier transform of the coordinate space function.
A great advantage in the correlated Gaussian function $F$ is that its 
Fourier transform is a linear combination of the 
correlated Gaussian functions in the momentum space. Thus 
by expressing ${\vi p}_i-{\vi p}_j$ or ${\vi p}_i-\frac{1}{N}{\vi \pi}_N$ as 
$\widetilde{\zeta}{\vi \pi}$, we can calculate 
the matrix element of the momentum-dependent 
operators in exactly the same way as in the coordinate
space. 

As in the case with two global vectors~\cite{DGVR}, 
the transformation from the coordinate to momentum space is achieved 
by a function 
\begin{equation}
\Phi({\vi k},{\vi x})=\frac{1}{(2\pi)^{\frac{3}{2}(N-1)}}\, 
\exp\,(i\tilde{\vi k}{\vi x}),
\end{equation}
where ${\vi k}$ is an $(N\!-\!1)$-dimensional column 
vector whose $i$th element is ${\vi k}_i$. With a straightforward integration 
together with the recoupling of angular momenta, we can show that 
\begin{eqnarray}
& &\langle \Phi({\vi k},{\vi x})\vert 
F_{L_1 L_2 (L_{12})L_3  L M}(u_1,u_2,u_3,A,{\vi x})\rangle
\nonumber \\
& &\quad =\frac{(-i)^{L_1+L_2+L_3}} {({\rm det}A)^{3/2}}
 \sum_{\ell_1 \ell_2 \ell_3 \ell_{12}}
{\cal K}(L_1L_2(L_{12})L_3 L; \ell_1 \ell_2 \ell_3 \ell_{12})
\nonumber \\
& &\quad \times F_{L_1\!-\!\ell_1\!-\!\ell_2\ L_2\!-\!\ell_1\!-\!\ell_3\ 
(\ell_{12})\ L_3\!-\!\ell_2\!-\!\ell_3\  L M}
(A^{-1}u_1, A^{-1}u_2, A^{-1}u_3, A^{-1},{\vi k}),
\end{eqnarray}
where the coefficient ${\cal K}$ is given by
\begin{eqnarray}
& &{\cal K}(L_1L_2(L_{12})L_3 L; \ell_1 \ell_2 \ell_3 \ell_{12})
\nonumber \\
& &\ =\frac{(-1)^{L-L_3+\ell_2+\ell_3-\ell_{12}}}
{\sqrt{2L\!+\!1}}
\frac{B_{L_1}B_{L_2}B_{L_3}}
{B_{\ell_1}B_{\ell_2}B_{\ell_3}B_{L_1\!-\!\ell_1\!-\!\ell_2}
B_{L_2\!-\!\ell_1\!-\!\ell_3}B_{L_3\!-\!\ell_2\!-\!\ell_3}}
\nonumber \\
& &\quad \times 
\sqrt{(2\ell_1\!+\!1)(2\ell_2\!+\!1)(2\ell_3\!+\!1)
(2(L_1\!-\!\ell_1\!-\!\ell_2)+1)(2(L_2\!-\!\ell_1\!-\!\ell_3)+1)(2(L_3\!-\!\ell_2\!-\!\ell_3)+1)}
\nonumber \\
& &\quad \times  X(\ell_1 \ell_2 \ell_3)
 Z(L_1\!-\!\ell_1\!-\!\ell_2\ \ell_1\!+\!\ell_2)
 Z(L_2\!-\!\ell_1\!-\!\ell_3\ \ell_1\!+\!\ell_3)
 Z(L_3\!-\!\ell_2\!-\!\ell_3\ \ell_2\!+\!\ell_3)
\nonumber \\
& &\quad \times U(\ell_{12}\ L_3\!-\!\ell_2\!-\!\ell_3\ L_{12}\ L_3; L\
 \ell_2\!+\!\ell_3)
\left[
\begin{array}{ccc}
L_1\!-\!\ell_1\!-\!\ell_2 & L_1 & \ell_1\!+\!\ell_2\\
L_2\!-\!\ell_1\!-\!\ell_3 & L_2 & \ell_1\!+\!\ell_3\\
\ell_{12} & L_{12} & \ell_2\!+\!\ell_3 \\
\end{array}
\right]
\nonumber \\
& &\quad \times 
(\widetilde{u_1}A^{-1}u_2)^{\ell_1}\ 
(\widetilde{u_1}A^{-1}u_3)^{\ell_2}\ 
(\widetilde{u_2}A^{-1}u_3)^{\ell_3},
\end{eqnarray}
where $Z$ and $X$ are defined in Appendix A. 
Non-negative integers $\ell_i$ run over all possible values that satisfy 
$\ell_1\!+\!\ell_2 \leq L_1,\ \ell_1\!+\!\ell_3 \leq L_2,\
\ell_2\!+\!\ell_3 \leq L_3$. The value of ${\ell}_{12}$ is restricted by
the triangular relations among 
($\ell_{12}, L_1\!-\!\ell_1\!-\!\ell_2, L_2\!-\!\ell_1\!-\!\ell_3$) 
and ($\ell_{12}, L_{12}, \ell_2\!+\!\ell_3 $).

\end{document}